\documentclass[iop,apj,numberedappendix,twocolappendix]{emulateapj}

\usepackage{apjfonts}
\usepackage{natbib}
\usepackage{amssymb}
\usepackage{enumitem}
\usepackage[breaklinks,colorlinks,citecolor=blue,linkcolor=magenta]{hyperref}

\newcommand{\magiicat}{\hbox{{\rm MAG}{\sc ii}CAT}}
\newcommand{\MgIIdblt}{{\rm Mg}\kern 0.1em{\sc ii}~$\lambda\lambda 2796, 2803$}
\newcommand{\MgII}{\hbox{{\rm Mg}\kern 0.1em{\sc ii}}}
\newcommand{\OII}{\hbox{[{\rm O}\kern 0.1em{\sc ii}]}}
\newcommand{\kms}{\hbox{km~s$^{-1}$}}
\newcommand{\etal}{et~al.}
\newcommand{\vfifty}{\hbox{$\Delta v(50)$}}
\newcommand{\vninety}{\hbox{$\Delta v(90)$}}
\newcommand{\blue}{\hbox{$B-K<1.4$}}
\newcommand{\red}{\hbox{$B-K\geq1.4$}}
\newcommand{\lowz}{\hbox{$z_{\rm gal}<0.656$}}
\newcommand{\highz}{\hbox{$z_{\rm gal}\geq0.656$}}
\newcommand{\lowdr}{\hbox{$D/R_{\rm vir}<0.24$}}
\newcommand{\highdr}{\hbox{$D/R_{\rm vir}\geq0.24$}}

\shorttitle{\sc ~{\magiicat} Kinematics: Quiescent Evolution}
\shortauthors{\sc Nielsen {\etal}}
\slugcomment{Accepted by ApJ, January 8, 2016}

\begin{document}

\title{~{\magiicat} IV. Kinematics of the Circumgalactic Medium and
  Evidence for Quiescent Evolution Around Red Galaxies}

\author{
Nikole M. Nielsen$^{1,2}$,
Christopher W. Churchill$^{2}$,
Glenn G. Kacprzak$^{1}$,
Michael T. Murphy$^{1}$,
and
Jessica L. Evans$^{2}$
}

\affil{$^1$ Centre for Astrophysics and Supercomputing, Swinburne
  University of Technology, Hawthorn, Victoria 3122, Australia;
  nikolenielsen@swin.edu.au\\
  $^2$ Department of Astronomy, New Mexico State University, Las
  Cruces, NM 88003, USA}

\begin{abstract}

The equivalent widths of {\MgII} absorption in the circumgalactic
medium (CGM) trace the global star formation rate up to $z<6$, are
larger for star-forming galaxies than passively-evolving galaxies, and
decrease with increasing distance from the galaxy. We delve further
into the physics involved by investigating gas kinematics and cloud
column density distributions as a function of galaxy color, redshift,
and projected distance from the galaxy (normalized by galaxy virial
radius, $D/R_{\rm vir}$). For 39 isolated galaxies at $0.3<z_{\rm
  gal}<1.0$, we have detected {\MgII} absorption in high-resolution
($\Delta v\simeq 6.6$~{\kms}) spectra of background quasars within a
projected distance of $7<D<190$~kpc. We characterize the absorption
velocity spread using pixel-velocity two-point correlation
functions. Velocity dispersions and cloud column densities for blue
galaxies do not differ with redshift nor with $D/R_{\rm vir}$. This
suggests that outflows continually replenish the CGM of blue galaxies
with high velocity dispersion, large column density gas out to large
distances. Conversely, absorption hosted by red galaxies evolves with
redshift where the velocity dispersions (column densities) are smaller
(larger) at {\lowz}. After taking into account larger possible
velocities in more massive galaxies, we find that there is no
difference in the velocity dispersions or column densities for
absorption hosted by red galaxies with $D/R_{\rm vir}$. Thus, a lack
of outflows in red galaxies causes the CGM to become more quiescent
over time, with lower velocity dispersions and larger column densities
towards lower $z_{\rm gal}$. The quenching of star formation appears
to affect the CGM out to $D/R_{\rm vir}=0.75$.

\end{abstract}


\section{Introduction}
\label{sec:intro}

Through extensive observations and detailed simulations, it has become
clear that the baryon cycle plays a key role in governing the
evolution of galaxies \citep[e.g.,][]{oppenheimer08,
  lilly-bathtub}. In this scenario, galaxies grow by accreting
pristine gas from the intergalactic medium (IGM), which is the fuel
for star formation. Intense star formation and/or supernovae then
drive outflowing galactic-scale winds, entraining portions of the
interstellar medium (ISM) and coplanar gas along the way. Serving as
the interface between the IGM and the ISM, the circumgalactic medium
(CGM) contains the gas that has yet to accrete onto the galaxy itself
and stores outflowing material until it re-accretes onto the
galaxy. The CGM is also massive, with estimates of $M_{\rm
  CGM}>10^9M_{\odot}$ for $\sim L_{\ast}$ galaxies \citep{thom11,
  tumlinson11, werk13}, a gas mass comparable to the gas in galaxies
themselves (i.e., the ISM).

The CGM has been studied extensively by using quasar absorption lines
in which a background quasar sightline pierces the CGM within a few
hundred kiloparsecs projected on the sky. Much of this work has been
focused on {\MgIIdblt} absorption \citep[e.g.,][]{bb91, sdp94,
  guillemin97, steidel97, bc09, chen10a, kcems11, lan14}, which is
observable from the ground at optical wavelengths over a large
redshift range ($0.1 < z < 2.5$). {\MgII} has been found to trace two
critical components of the baryon cycle: accretion
\citep[e.g.,][]{steidel02, ggk-sims, ribaudo11, ggk1317, martin12,
  rubin-accretion, bouche13, crighton13} and galactic-scale outflowing
winds.  \citep[e.g.,][]{weiner09, rubin-winds, rubin-winds14,
  bouche12, martin12, bordoloi14-model, bordoloi14, kacprzak14}.

Determining which aspect of the baryon cycle {\MgII} absorbers trace
has thus far required examining the velocities of the gas with respect
to the galaxy systemic velocity. For example, the signatures of
accreting and/or rotating gas include absorption that is located to
one side (i.e., entirely bluewards or redwards) of the galaxy systemic
velocity \citep{steidel02, ggk-sims, stewart11, bouche13} or observing
redshifted absorption with respect to the galaxy systemic velocity in
a ``down-the-barrel'' approach \citep{martin12,
  rubin-accretion}. Outflows are commonly observed as blueshifted
absorption (with respect to the galaxy systemic velocity) for the same
down-the-barrel approach \citep[e.g,][]{weiner09, rubin-winds,
  rubin-winds14, martin12, bordoloi14}. Recently, \citet{fox15} found
a quasar whose sightline passes through the ``Fermi bubbles'' located
near the center of the Milky Way Galaxy. In low ionization absorption,
they found velocity structure in the form of smaller column density,
higher velocity components that were consistent with the front and
back sides of the Fermi bubbles. This result provides hints that the
velocity structure of the absorbers themselves, rather than just the
velocity with respect to the galaxy, is dependent on baryon cycle
processes.

Other works have alluded to the kinematic properties of the absorption
itself in the presence of outflows by examining only the {\MgII}
equivalent width, $W_r(2796)$. This method finds that $W_r(2796)$ is
dependent on galaxy color (star formation rate), azimuthal angle,
and/or inclination, where larger values tend to be associated with
blue galaxies and sightlines probing galaxies near their minor axes
\citep{bordoloi11, bordoloi14-model, kcems11, kcn12, bouche12}, both
of which are known to host bipolar outflows. Given that $W_r(2796)$
correlates with the number of clouds or Voigt profile components
\citep{pb90, cvc03, evans-thesis}, this indicates that either the
column densities, velocity spreads, or both are larger in the presence
of outflows. The kinematics and column densities may also differ when
associated with accretion, though accreting gas is harder to detect
due to its small covering fraction \citep[$\sim 6\%$;][]{martin12},
and the fact that outflowing gas dominates the absorption signal.

Using $\sim 8500$ strong {\MgII} absorbers ($0.7 < W_r(2796) <
6.0$~{\AA}) at $0.4 < z < 1.3$, \citet{menard11} found a $15\sigma$
correlation between $W_r(2796)$ and the {\OII} luminosity of the
associated galaxy, which provides an estimate for the star formation
rate. With this correlation, the authors were able to show that the
star formation rate probed by strong {\MgII} absorption follows the
star formation history up to at least $z\sim 2$. \citet{matejek12}
used similar methods for {\MgII} absorption in infrared wavelengths
with the FIRE spectrograph on Magellan and were able to extend the
redshift range out to $z<6$. They found that the star formation rate
as probed by {\MgII} traces the global star formation rate out to
$z<6$, including the peak at $z\sim 2-3$. Thus, the {\MgII} equivalent
width, and possibly the velocity/column density structure, of the
strongest absorbers traces the global star formation rate up to $z=6$.

Additionally, the equivalent width of {\MgII} absorption has long been
found to anti-correlate with impact parameter at up to a $7.9\sigma$
significance \citep[e.g.,][]{lanzetta90, sdp94, ggk08, chen10a,
  magiicat2, magiicat1}, where the equivalent width decreases with
increasing impact parameter. More recently, \citet{cwc-masses} found
that $W_r(2796)$ anti-correlates with the impact parameter normalized
by the virial radius of the galaxy, $D/R_{\rm vir}$, at the $\sim
9\sigma$ level, a more significant anti-correlation than with $D$
alone. Given that the equivalent width is proportional to the number
of clouds (or velocity components) fit with Voigt profile modeling
\citep[e.g.,][]{pb90, cvc03, evans-thesis}, this anti-correlation may
be due to the column densities, velocity spreads, or both diminishing
with projected distance from the galaxy.

These results indicate that examining the kinematic structure and/or
column density distribution of the gas traced by {\MgII} absorption
over time and space is critical in understanding the detailed physics
of the baryon cycle processes occurring in the CGM. In particular,
studying the detailed velocity structure and column density
distributions of the absorbers constrains the gas physics involved.

Many works examining the detailed {\MgII} absorber kinematics have
focused on the clustering of VP components in {\MgII} absorbers by
constructing a two-point correlation function (TPCF) for their samples
using VP component velocities \citep{ssb88, pb90, cwc-thesis, cv01,
  cvc03, evans-thesis}. \citet{cvc03} fitted their TPCFs with two
Gaussian components, where the narrower component is associated with
vertical dispersions in face-on galaxy disks. The second and more
broad component may represent the rotational motions in edge-on disks
\citep{cvc03}. More recently, \citet{evans-thesis} required three
components to fit their TPCF because their sample was more than an
order of magnitude larger than previous works and therefore more
sensitive to an extended tail in the
distribution. \citet{evans-thesis} did not try to interpret their
fitted Gaussian components, stating that doing so would be an
oversimplification. This is reasonable since their absorber sample
spans a large redshift range ($0.1 < z_{\rm abs} < 2.6$) and likely
probes the CGM of a variety of galaxy types.

What these previous absorber kinematics studies lack is the connection
between the detailed kinematics of these absorbers to the properties
of their host galaxies. In a companion paper \citep[Paper V of the
  {\magiicat} series;][]{magiicat5}, we examined the kinematics as a
function of galaxy color, inclination, and the azimuthal angle at
which the CGM is probed. We characterized the kinematics by creating
pixel-velocity two-point correlation functions (TPCFs; similar to the
TPCFs used in previous works) for various color and orientation
subsamples. We found that absorbers hosted by blue galaxies in
``face-on'' orientations, especially near the projected galaxy minor
axis, have the largest velocity dispersions, while absorbers hosted by
red galaxies for all orientations have small velocity dispersions. We
concluded that for blue galaxies, gas entrained in bipolar outflows
may have large velocity dispersions and may be fragmented into clouds
with smaller column densities, while gas accreting onto or rotating
around the galaxy along the major axis (especially for ``edge-on''
orientations) may be more coherent, due to larger cloud column
densities and smaller velocity dispersions. Conversely, we attributed
small velocity dispersions for red galaxies along the minor axis to a
lack of outflows, but larger velocity dispersions along the major axis
may indicate gas that is accreting onto or rotating around the galaxy.

In this paper, we examine the kinematics of {\MgII} absorption by
using the same pixel-velocity TPCF method as \citet{magiicat5}, but do
so as a function of galaxy rest-frame $B-K$ color, redshift, $z_{\rm
  gal}$, and impact parameter normalized by the virial radius,
$D/R_{\rm vir}$. We organize this paper as
follows. Section~\ref{sec:methods} details our sample, including both
galaxy properties and quasar spectra. Section~\ref{sec:kinematics}
briefly characterizes quasar absorption line kinematics in terms of
Voigt profile components and then details our methods for calculating
pixel-velocity two-point velocity correlation functions, presenting
only a bivariate analysis of the TPCFs with galaxy rest-frame color,
$B-K$. Section~\ref{sec:results} presents a multivariate analysis in
the TPCFs for cuts in galaxy color, redshift, and $D/R_{\rm vir}$. We
discuss our results in Section~\ref{sec:discussion}, and summarize and
conclude our findings in Section~\ref{sec:conclusions}.

\begin{deluxetable*}{lllccccccccc}
\tabletypesize{\footnotesize}
\tablecolumns{12}
\tablewidth{0pt}
\setlength{\tabcolsep}{0.06in}
\tablecaption{Absorber--Galaxy Properties \label{tab:props}}
\tablehead{
  \colhead{(1)}                      &
  \colhead{(2)}                      &
  \colhead{(3)}                      &
  \colhead{(4)}                      &
  \colhead{(5)}                      &
  \colhead{(6)}                      &
  \colhead{(7)}                      &
  \colhead{(8)}                      &
  \colhead{(9)}                      &
  \colhead{(10)}                     &
  \colhead{(11)}                     &
  \colhead{(12)}                     \\
  \colhead{QSO}                      &
  \colhead{J-Name}                   &
  \colhead{$z_{\rm gal}$}             &
  \colhead{$B-K$}                    &
  \colhead{$D$}                      &
  \colhead{$D/R_{\rm vir}$}           &
  \colhead{$\log (M_{\rm h}/M_{\odot})$} &
  \colhead{$V_{\rm circ}$}            &
  \colhead{$z_{\rm abs}$}             &
  \colhead{$W_r(2796)$}              &
  \colhead{$\log~N({\MgII})$}        &
  \colhead{Ref\tablenotemark{a}}     \\
  \multicolumn{4}{c}{}               &
  \colhead{(kpc)}                    &
  \colhead{}                         &
  \colhead{}                         &
  \colhead{({\kms})}                 &
  \colhead{}                         &
  \colhead{(\AA)}                    &
  \colhead{(cm$^{-2}$)}               &
  \colhead{}                         }    
\startdata

$0002+051$ & J$000520.21+052411.80 $ & $0.298   $ & $2.43   $ & $  59.2$ & $0.31$ & $ 12.0_{-0.2}^{+0.3} $ & $ 211_{-26}^{+45} $ & $0.298059$ & $0.244\pm0.003$ & $13.14\pm0.08$ & 1  \\[3pt]
$0002+051$ & J$000520.21+052411.80 $ & $0.592   $ & $2.05   $ & $  36.0$ & $0.14$ & $ 12.3_{-0.2}^{+0.2} $ & $ 291_{-29}^{+38} $ & $0.591365$ & $0.102\pm0.002$ & $12.60\pm0.11$ & 1  \\[3pt]
$0002+051$ & J$000520.21+052411.80 $ & $0.85180 $ & $0.74   $ & $  25.9$ & $0.14$ & $ 11.8_{-0.2}^{+0.2} $ & $ 220_{-24}^{+40} $ & $0.851393$ & $1.089\pm0.008$ & $14.43\pm0.24$ & 1  \\[3pt]
$0058+019$ & J$010054.15+021136.52 $ & $0.6128  $ & $1.32   $ & $  29.5$ & $0.24$ & $ 11.4_{-0.2}^{+0.4} $ & $ 151_{-20}^{+51} $ & $0.612586$ & $1.684\pm0.004$ & $15.74\pm0.12$ & 2  \\[3pt]
$0102-190$ & J$010516.82-184641.9  $ & $1.025   $ & $\cdots $ & $  40.0$ & $0.17$ & $ 12.1_{-0.1}^{+0.1} $ & $ 284_{-25}^{+31} $ & $1.026450$ & $0.946\pm0.010$ & $15.15\pm0.45$ & 3  \\[3pt]
$0117+213$ & J$012017.20+213346.00 $ & $0.5763  $ & $2.09   $ & $   7.8$ & $0.02$ & $ 12.9_{-0.1}^{+0.1} $ & $ 415_{-37}^{+35} $ & $0.576398$ & $0.902\pm0.007$ & $\sim15.31$\tablenotemark{b} & 2  \\[3pt]
$0117+213$ & J$012017.20+213346.00 $ & $0.729   $ & $2.12   $ & $  55.4$ & $0.14$ & $ 12.9_{-0.1}^{+0.1} $ & $ 434_{-35}^{+33} $ & $0.729077$ & $0.244\pm0.005$ & $13.04\pm0.08$ & 1  \\[3pt]
$0150-202$ & J$015227.32-200107.10 $ & $0.780   $ & $1.03   $ & $  54.7$ & $0.26$ & $ 12.1_{-0.2}^{+0.2} $ & $ 252_{-27}^{+38} $ & $0.779796$ & $0.404\pm0.016$ & $15.80\pm0.17$ & 3  \\[3pt]
$0229+131$ & J$023145.89+132254.71 $ & $0.4167  $ & $2.04   $ & $  36.9$ & $0.14$ & $ 12.4_{-0.2}^{+0.2} $ & $ 285_{-29}^{+34} $ & $0.417338$ & $0.816\pm0.020$ & $13.83\pm0.22$ & 1  \\[3pt]
$0235+164$ & J$023838.93+163659.27 $ & $0.852   $ & $1.48   $ & $   7.6$ & $0.02$ & $ 12.6_{-0.1}^{+0.1} $ & $ 370_{-32}^{+31} $ & $0.852255$ & $0.505\pm0.004$ & $13.68\pm0.12$ & 3  \\[3pt]
$0302-223$ & J$030450.10-221157.00 $ & $0.418   $ & $\cdots $ & $ 126.0$ & $0.20$ & $ 13.5_{-0.1}^{+0.1} $ & $ 625_{-52}^{+47} $ & $0.420411$ & $0.727\pm0.028$ & $14.76\pm0.96$ & 3  \\[3pt]
$0302-223$ & J$030450.10-221157.00 $ & $1.000   $ & $\cdots $ & $  61.2$ & $0.31$ & $ 12.0_{-0.1}^{+0.2} $ & $ 248_{-24}^{+34} $ & $1.009382$ & $1.099\pm0.036$ & $15.22\pm0.50$ & 2  \\[3pt]
$0334-204$ & J$033626.90-201940.00 $ & $1.120   $ & $\cdots $ & $  64.3$ & $0.19$ & $ 12.6_{-0.1}^{+0.1} $ & $ 404_{-32}^{+30} $ & $1.117706$ & $1.706\pm0.020$ & $16.85\pm0.30$ & 3  \\[3pt]
$0349-146$ & J$035128.54-142908.71 $ & $0.3567  $ & $0.28   $ & $  71.3$ & $0.42$ & $ 11.9_{-0.2}^{+0.3} $ & $ 193_{-25}^{+52} $ & $0.357168$ & $0.175\pm0.007$ & $13.86\pm0.30$ & 1  \\[3pt]
$0454-220$ & J$045608.92-215909.40 $ & $0.48382 $ & $1.66   $ & $ 107.1$ & $0.44$ & $ 12.3_{-0.2}^{+0.2} $ & $ 270_{-28}^{+38} $ & $0.483337$ & $0.426\pm0.007$ & $13.68\pm0.39$ & 1  \\[3pt]
$0454+039$ & J$045647.17+040052.94 $ & $0.8596  $ & $\cdots $ & $  16.0$ & $0.14$ & $ 11.2_{-0.2}^{+0.4} $ & $ 145_{-19}^{+49} $ & $0.859569$ & $1.476\pm0.009$ & $\sim15.51$\tablenotemark{b} & 2  \\[3pt]
$0827+243$ & J$083052.08+241059.82 $ & $0.5247  $ & $2.23   $ & $  37.2$ & $0.15$ & $ 12.3_{-0.2}^{+0.2} $ & $ 282_{-29}^{+38} $ & $0.524966$ & $2.419\pm0.012$ & $\sim15.19$\tablenotemark{b} & 1  \\[3pt]
$0836+113$ & J$083933.01+111203.82 $ & $0.78682 $ & $0.86   $ & $  26.8$ & $0.15$ & $ 11.8_{-0.2}^{+0.3} $ & $ 212_{-24}^{+46} $ & $0.786725$ & $2.113\pm0.019$ & $15.54\pm7.39$ & 1  \\[3pt]
$1019+309$ & J$102230.29+304105.11 $ & $0.346   $ & $1.23   $ & $  46.0$ & $0.27$ & $ 11.9_{-0.2}^{+0.3} $ & $ 193_{-25}^{+52} $ & $0.346246$ & $0.628\pm0.017$ & $15.54\pm0.41$ & 3  \\[3pt]
$1038+064$ & J$104117.16+061016.92 $ & $0.4432  $ & $2.81   $ & $  55.9$ & $0.29$ & $ 12.0_{-0.2}^{+0.3} $ & $ 221_{-26}^{+43} $ & $0.441453$ & $0.673\pm0.011$ & $13.72\pm0.26$ & 1  \\[3pt]
$1100-264$ & J$110325.29-264515.7  $ & $0.359   $ & $\cdots $ & $  60.8$ & $0.31$ & $ 12.0_{-0.2}^{+0.3} $ & $ 216_{-27}^{+46} $ & $0.358989$ & $0.545\pm0.001$ & $14.26\pm0.08$ & 3  \\[3pt]
$1148+387$ & J$115129.37+382552.35 $ & $0.5536  $ & $1.19   $ & $  20.4$ & $0.11$ & $ 12.0_{-0.2}^{+0.3} $ & $ 224_{-27}^{+45} $ & $0.553363$ & $0.640\pm0.013$ & $13.47\pm0.13$ & 1  \\[3pt]
$1209+107$ & J$121140.59+103002.02 $ & $0.392   $ & $1.02   $ & $  37.5$ & $0.27$ & $ 11.6_{-0.2}^{+0.4} $ & $ 158_{-22}^{+58} $ & $0.392924$ & $1.187\pm0.005$ & $13.94\pm0.30$ & 1  \\[3pt]
$1222+228$ & J$122527.39+223513.0  $ & $0.5502  $ & $2.17   $ & $  37.7$ & $0.26$ & $ 11.6_{-0.2}^{+0.4} $ & $ 170_{-23}^{+54} $ & $0.550198$ & $0.094\pm0.009$ & $12.45\pm0.36$ & 1  \\[3pt]
$1229-021$ & J$123200.01-022405.27 $ & $0.7546  $ & $1.33   $ & $  12.4$ & $0.07$ & $ 11.8_{-0.2}^{+0.3} $ & $ 215_{-24}^{+43} $ & $0.756903$ & $0.303\pm0.003$ & $13.44\pm0.07$ & 2  \\[3pt]
$1241+176$ & J$124410.82+172104.52 $ & $0.550   $ & $1.34   $ & $  21.1$ & $0.12$ & $ 11.8_{-0.2}^{+0.3} $ & $ 202_{-25}^{+47} $ & $0.550482$ & $0.465\pm0.011$ & $13.63\pm0.12$ & 1  \\[3pt]
$1246-057$ & J$124913.85-055919.07 $ & $0.637   $ & $1.63   $ & $  29.0$ & $0.18$ & $ 11.7_{-0.2}^{+0.3} $ & $ 192_{-23}^{+45} $ & $0.639909$ & $0.450\pm0.004$ & $13.74\pm0.26$ & 1  \\[3pt]
$1248+401$ & J$125048.32+395139.48 $ & $0.7725  $ & $1.28   $ & $  35.4$ & $0.23$ & $ 11.6_{-0.2}^{+0.3} $ & $ 185_{-23}^{+48} $ & $0.772957$ & $0.695\pm0.005$ & $13.85\pm2.32$ & 2  \\[3pt]
$1254+047$ & J$125659.92+042734.39 $ & $0.9341  $ & $1.22   $ & $  12.5$ & $0.09$ & $ 11.6_{-0.2}^{+0.3} $ & $ 184_{-22}^{+47} $ & $0.934231$ & $0.338\pm0.005$ & $13.25\pm0.10$ & 2  \\[3pt]
$1317+277$ & J$131956.23+272808.22 $ & $0.6610  $ & $1.45   $ & $ 103.1$ & $0.46$ & $ 12.1_{-0.2}^{+0.2} $ & $ 259_{-27}^{+37} $ & $0.660049$ & $0.320\pm0.006$ & $13.13\pm0.43$ & 1  \\[3pt]
$1331+170$ & J$133335.78+164904.01 $ & $0.7443  $ & $2.02   $ & $  30.5$ & $0.15$ & $ 12.0_{-0.2}^{+0.2} $ & $ 245_{-27}^{+39} $ & $0.744642$ & $1.836\pm0.003$ & $14.19\pm0.10$ & 2  \\[3pt]
$1354+195$ & J$135704.43+191907.37 $ & $0.4592  $ & $1.40   $ & $  45.1$ & $0.28$ & $ 11.7_{-0.2}^{+0.3} $ & $ 184_{-24}^{+48} $ & $0.456598$ & $0.773\pm0.015$ & $13.90\pm0.92$ & 1  \\[3pt]
$1424-118$ & J$142738.10-120350.00 $ & $0.3404  $ & $1.77   $ & $  85.9$ & $0.46$ & $ 12.0_{-0.2}^{+0.3} $ & $ 209_{-27}^{+48} $ & $0.341716$ & $0.100\pm0.015$ & $12.61\pm0.06$ & 1  \\[3pt]
$1548+092$ & J$155103.39+090849.25 $ & $0.7703  $ & $0.68   $ & $  40.5$ & $0.33$ & $ 11.4_{-0.2}^{+0.4} $ & $ 155_{-20}^{+53} $ & $0.770643$ & $0.229\pm0.018$ & $13.26\pm0.02$ & 3  \\[3pt]
 SDSS      & J$160726.77+471251.37 $ & $0.4980  $ & $1.41   $ & $ 188.6$ & $0.75$ & $ 12.3_{-0.2}^{+0.2} $ & $ 281_{-28}^{+38} $ & $0.497479$ & $1.237\pm0.037$ & $15.90\pm3.53$ & 3  \\[3pt]
$1622+238$ & J$162439.08+234512.20 $ & $0.3181  $ & $2.85   $ & $  54.4$ & $0.28$ & $ 12.0_{-0.2}^{+0.3} $ & $ 215_{-26}^{+45} $ & $0.317597$ & $0.491\pm0.010$ & $13.88\pm1.17$ & 1  \\[3pt]
$1622+238$ & J$162439.08+234512.20 $ & $0.4720  $ & $0.92   $ & $  34.0$ & $0.28$ & $ 11.4_{-0.2}^{+0.5} $ & $ 142_{-19}^{+54} $ & $0.471930$ & $0.769\pm0.006$ & $14.48\pm2.16$ & 1  \\[3pt]
$1622+238$ & J$162439.08+234512.20 $ & $0.6560  $ & $0.93   $ & $  99.3$ & $0.69$ & $ 11.6_{-0.2}^{+0.4} $ & $ 173_{-22}^{+48} $ & $0.656106$ & $1.446\pm0.006$ & $\sim14.82$\tablenotemark{b} & 1  \\[3pt]
$1622+238$ & J$162439.08+234512.20 $ & $0.7975  $ & $1.66   $ & $  71.3$ & $0.35$ & $ 12.0_{-0.2}^{+0.2} $ & $ 247_{-27}^{+40} $ & $0.797078$ & $0.468\pm0.008$ & $13.28\pm0.06$ & 1  \\[3pt]
$1622+238$ & J$162439.08+234512.20 $ & $0.8909  $ & $0.41   $ & $  23.2$ & $0.14$ & $ 11.7_{-0.2}^{+0.3} $ & $ 201_{-24}^{+43} $ & $0.891276$ & $1.548\pm0.004$ & $\sim14.90$\tablenotemark{b} & 1  \\[3pt]
$2128-123$ & J$213135.26-120704.79 $ & $0.430   $ & $2.06   $ & $  48.1$ & $0.24$ & $ 12.0_{-0.2}^{+0.2} $ & $ 225_{-26}^{+43} $ & $0.429812$ & $0.452\pm0.008$ & $\sim14.18$\tablenotemark{b} & 2  \\[3pt]
$2145+067$ & J$214805.45+065738.60 $ & $0.790   $ & $1.39   $ & $  40.8$ & $0.19$ & $ 12.1_{-0.2}^{+0.2} $ & $ 256_{-27}^{+39} $ & $0.790866$ & $0.547\pm0.005$ & $13.42\pm0.62$ & 2  \\[3pt]
$2206-199$ & J$220852.07-194359.0  $ & $0.752   $ & $\cdots $ & $  11.7$ & $0.06$ & $ 11.9_{-0.2}^{+0.3} $ & $ 221_{-25}^{+43} $ & $0.751923$ & $0.890\pm0.002$ & $16.23\pm0.04$ & 2  \\[3pt]
$2206-199$ & J$220852.07-194359.0  $ & $0.948   $ & $0.74   $ & $  86.9$ & $0.37$ & $ 12.2_{-0.1}^{+0.2} $ & $ 286_{-27}^{+35} $ & $0.948362$ & $0.256\pm0.003$ & $13.18\pm0.07$ & 2  \\[3pt]
$2206-199$ & J$220852.07-194359.0  $ & $1.01655 $ & $0.63   $ & $ 104.4$ & $0.31$ & $ 12.6_{-0.1}^{+0.1} $ & $ 399_{-32}^{+30} $ & $1.017050$ & $1.058\pm0.004$ & $14.43\pm0.11$ & 2  \\[3pt]
$2231-002$ & J$223408.99+000001.69 $ & $0.8549  $ & $\cdots $ & $  23.6$ & $0.16$ & $ 11.6_{-0.2}^{+0.3} $ & $ 184_{-22}^{+45} $ & $0.855069$ & $0.784\pm0.004$ & $13.75\pm0.13$ & 2  \\[3pt]
$2343+125$ & J$234628.21+124859.9  $ & $0.7313  $ & $1.22   $ & $  32.5$ & $0.26$ & $ 11.4_{-0.2}^{+0.4} $ & $ 154_{-20}^{+53} $ & $0.731219$ & $1.655\pm0.006$ & $\sim16.21$\tablenotemark{b} & 2  \\[-5pt]

\enddata
\tablenotetext{a}{{\MgII} Absorption Measurements: (1)
  \citet{kcems11}, (2) \citet{evans-thesis}, and (3) This work.}
\tablenotetext{b}{At least one cloud is not well constrained,
  resulting in a large uncertainty.}
\end{deluxetable*}

\section{Sample and Data Analysis}
\label{sec:methods}

\subsection{Galaxy Properties}

All 39 galaxies ($0.3 < z_{\rm gal} < 1.0$) studied here are a subset
of the {\MgII} Absorber--Galaxy Catalog ({\magiicat}) and we refer the
reader to Paper I of the series \citep{magiicat1} for full details of
the data, the selection methods, and how galaxy properties were
determined. To summarize, each galaxy is spectroscopically identified
to be located at the redshift of an associated {\MgII} absorber
(whether absorption was detected and measured {\it a priori} or not)
and within a projected distance $D<200$~kpc from a background
quasar. All galaxies are isolated to the limits of the data available
\citep[for details, see][]{magiicat1}, where isolation is defined as
having no spectroscopically identified galaxy within 100~kpc
(projected) and within a line-of-sight velocity separation of
500~\kms. For each galaxy, we have spectroscopic redshifts, $z_{\rm
  gal}$, rest-frame $B$- and $K$-band AB magnitudes and luminosities,
rest-frame $B-K$ colors, and quasar--galaxy impact parameters, $D$. We
also have halo masses, $\log (M_{\rm h}/M_{\odot})$, virial radii,
$R_{\rm vir}$, and maximum circular velocities, $V_{\rm circ}$, from
halo abundance matching, the details for which are presented in Paper
III \citep{magiicat3}.

While the data for the absorber--galaxy pairs used here are published
elsewhere \citep{kcems11, evans-thesis, magiicat1, magiicat3}, we
present the galaxy and absorber data for each pair in
Table~\ref{tab:props}. Columns (1) and (2) are the quasar field names,
columns (3)--(8) are the galaxy properties and columns (9)--(12) are
the absorber properties. Columns (3)--(5) were published in
\citet{magiicat1}, while columns (7) and (8) and the $R_{\rm vir}$
values for column (6) were published in \citet{magiicat3}.

In order to investigate any dependencies of {\MgII} absorption on the
star formation rate over time, as well as any radial dependencies, we
slice our sample into various subsamples based on median galaxy
rest-frame color, $B-K$\footnote{This ensures the largest and (nearly)
  equal subsample sizes for comparison rather than cutting by the more
  traditional galaxy color bimodality (roughly 1.87 for this sample),
  which would result in small red subsamples and unequal subsample
  sizes.}, redshift, $z_{\rm gal}$, and impact parameter normalized by
the virial radius, $D/R_{\rm vir}$. The median value is appropriate
here as it allows for roughly equal sample sizes for comparison. A
summary is presented in Table~\ref{tab:v50}, which details the median
value(s) by which the subsamples are defined and the number of
galaxies in each subsample. The table also lists the median $z_{\rm
  gal}$ and $D/R_{\rm vir}$ for each subsample after the full sample
cuts are made.

\begin{figure}[t]
\includegraphics[width=\linewidth]{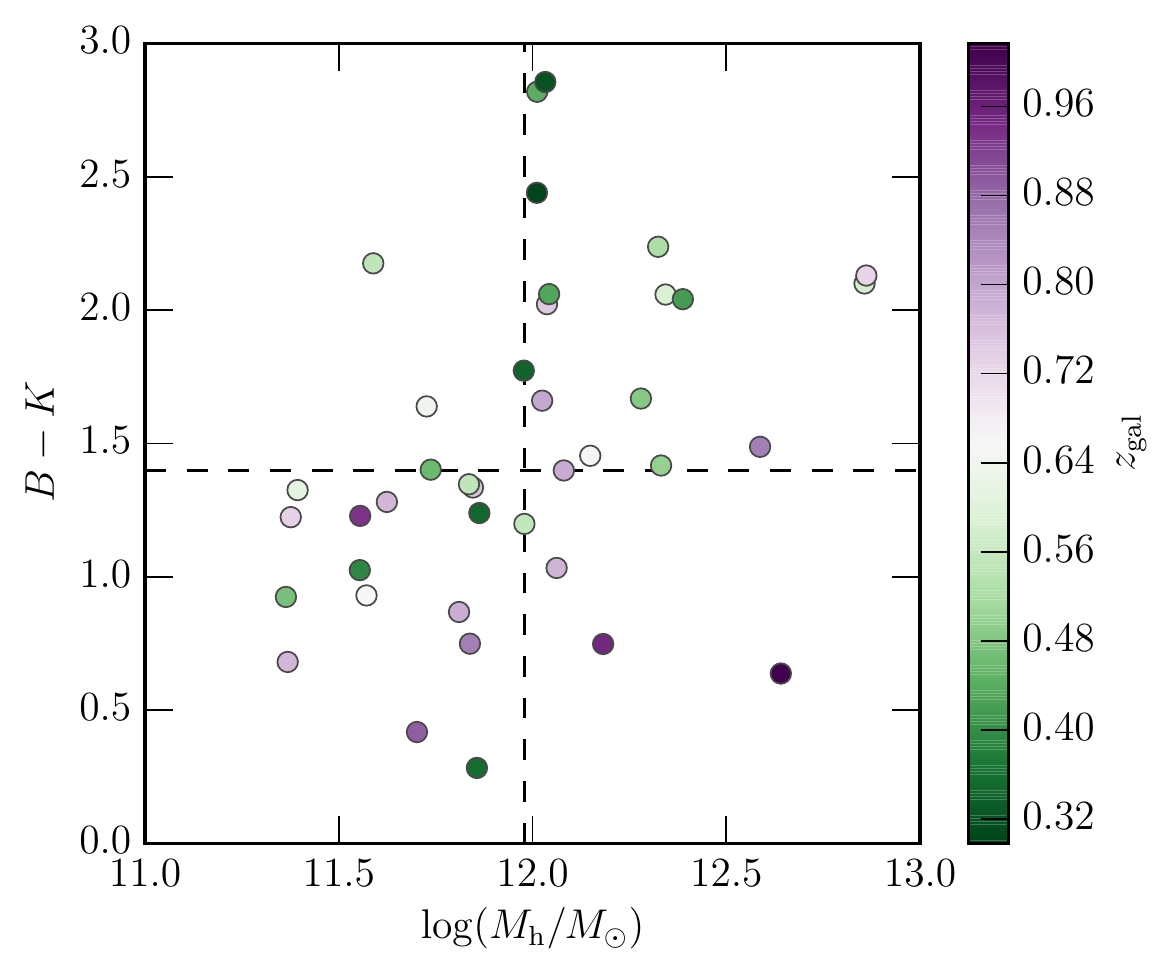}
\caption[]{Galaxy rest-frame $B-K$ color versus halo mass, $\log
  (M_{\rm h}/M_{\odot})$. Point colors indicate the galaxy redshift,
  $z_{\rm gal}$, where the median redshift of $\langle z_{\rm gal}
  \rangle = 0.656$ roughly corresponds to the white portion of the
  color bar. The vertical dashed line represents the median $\log
  (M_{\rm h}/M_{\odot})$ of the sample and the horizontal dashed line
  is the median $B-K$. A Kendall-$\tau$ rank correlation test
  comparing $B-K$ to $\log (M_{\rm h}/M_{\odot})$ found a correlation
  with a significance of $2.8\sigma$. This is represented as having
  most points located in the blue, low mass or red, high mass
  subsamples.}
\label{fig:BKMh}
\end{figure}

We note that, though the focus of this paper is on galaxy $B-K$
colors, it is difficult to disentangle effects due to color from those
due to galaxy halo masses, $\log (M_{\rm h}/M_{\odot})$. A
Kendall-$\tau$ rank correlation test on color and mass results in a
$2.8\sigma$ correlation such that redder galaxies tend to be more
massive. Figure~\ref{fig:BKMh} presents $B-K$ versus $\log (M_{\rm
  h}/M_{\odot})$ with points colored by $z_{\rm gal}$. Dashed lines
indicate the median color and mass of the sample. Almost all galaxies
in our sample lie within the blue, low mass or red, high mass regions
of Figure~\ref{fig:BKMh}, with the exception of four blue, high mass
galaxies and four red, low mass galaxies. If we instead conduct our
analysis with $\log (M_{\rm h}/M_{\odot})$, we find no significant
differences in the TPCFs when we compare blue samples to low mass
samples or red to high mass samples. Therefore, any differences we
find in the TPCFs is due to a color--mass dependence rather than just
a color dependence. To mitigate this, we account for the host galaxy
mass by normalizing velocities by the maximum circular velocity,
$V_{\rm circ}$, of the host galaxy.

We also examine possible trends between the properties we use to cut
the sample to rule out the possibility that any differences we may
find between subsamples are mainly due to biases in the data. We ran a
Kendall-$\tau$ rank correlation test between $B-K$ and $z_{\rm gal}$
and find an anti-correlation with a significance of $3.1\sigma$. In
this case, bluer galaxies tend to be located at higher redshift, as
can be seen in Figure~\ref{fig:BKMh}. Comparing $B-K$ and $D/R_{\rm
  vir}$ results in an insignificant anti-correlation at $0.8\sigma$,
while we also find an insignificant anti-correlation between $z_{\rm
  gal}$ and $D/R_{\rm vir}$ at $2.0\sigma$.

\subsection{Quasar Spectra}
\label{sec:spectra}

The sample we present here is a {\MgII} absorption-selected
sample. For each of the 39 isolated galaxies, we have a
high-resolution spectrum of a nearby background quasar in which
absorption is detected at the redshift of the galaxy. Quasar spectra
were observed with HIRES/Keck \citep{vogt-hires} or UVES/VLT
\citep{dekker-uves}. Most spectra and details of their reduction are
published in \citet{cwc-thesis}, \citet{cv01}, \citet{evans-thesis},
and/or \citet{kcems11}. We obtained two additional reduced HIRES/Keck
spectra through private communication with C.~C Steidel and
J.-R. Gauthier. These latter two spectra were reduced using the Mauna
Kea Echelle Extraction ({\sc
  makee}\footnote{http://www.astro.caltech.edu/\textasciitilde
  tb/makee/}) package.

Full explanations of how the {\MgII} absorption systems are identified
in the quasar spectra and Voigt profile fitted are presented in great
detail in \citet{cwc-thesis}, \citet{cv01}, \citet{cvc03},
\citet{evans-thesis}, and \citet{kcems11}. We present only a summary
of the process here.

\begin{figure*}[ht]
\includegraphics[width=\linewidth]{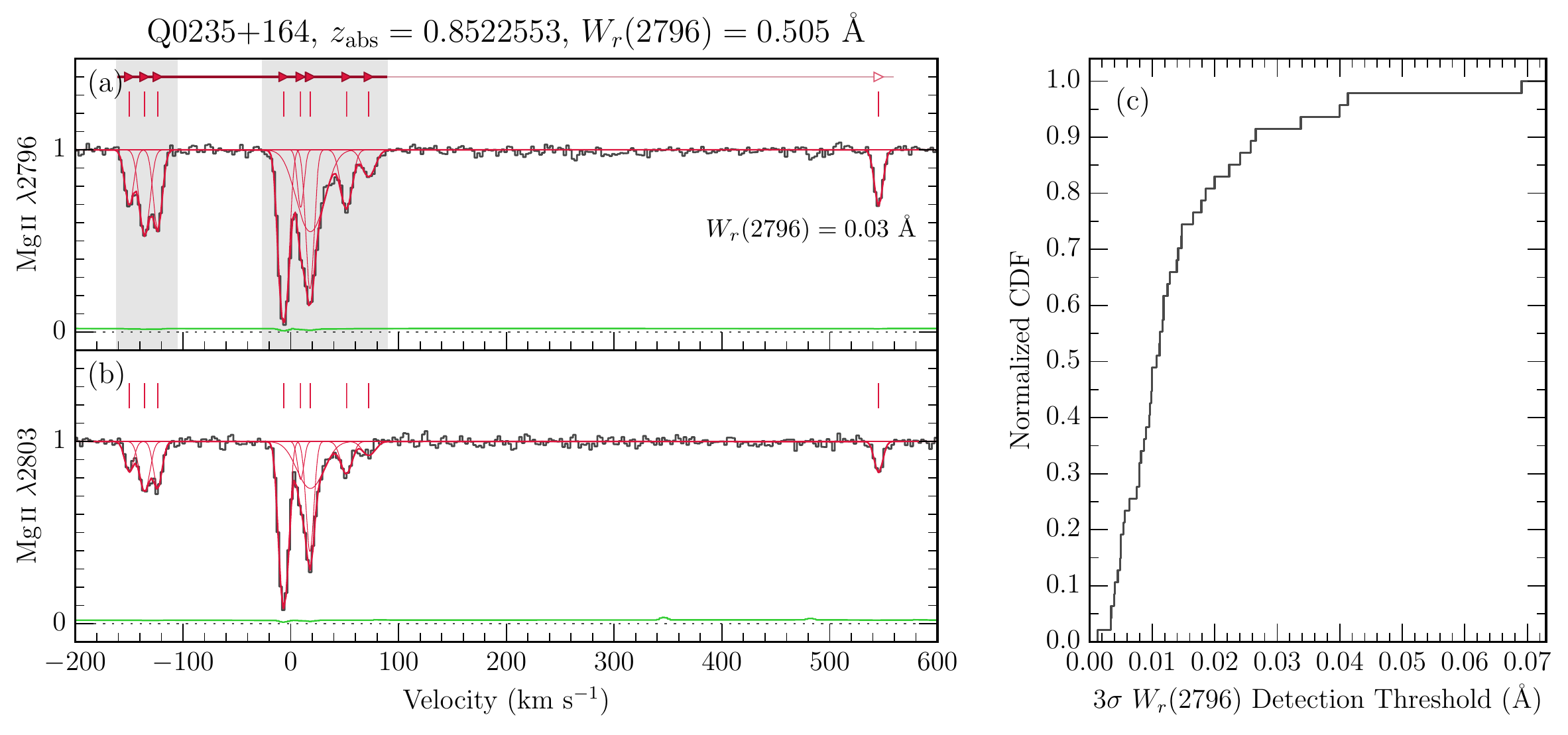}
\caption[]{An example quasar spectrum focused on the {\MgIIdblt}
  absorption doublet in panels (a) and (b). Black histograms are the
  data, thick red lines are the full Voigt profile fit to the data,
  and thin red lines are the individual Voigt profile components that
  compose the total fit. The VP components are centered on the
  vertical ticks at the top of each panel. The shaded regions indicate
  the pixels used in the TPCF analysis. The horizontal line and
  triangles at the top of panel (a) represent a simplified spectrum
  for use in Figures~\ref{fig:vvsEW} and \ref{fig:bivariate}(a). (c)
  Cumulative distribution function of the equivalent width detection
  threshold for the full kinematics sample. This sample is complete to
  $\sim95\%$ at 0.04~{\AA}. We do not include any kinematic subsystems
  for which $W_r(2796)<0.04$~{\AA}. An example subsystem below this
  limit is shown at $v\sim 550$~{\kms} in panel (a), which has
  $W_r(2796)=0.03$~{\AA}.}
\label{fig:specEW}
\end{figure*}

Using {\sc Sysanal} \citep{cwc-thesis, cv01, evans-thesis}, we detect
the {\MgIIdblt} absorption doublet in each spectrum with a $5\sigma$
($3\sigma$) significance criterion in the equivalent width spectrum
for the $\lambda 2796$ ($\lambda 2803$) line by following the
formalism of \citet{schneider93}. {\sc Sysanal} determines velocity or
wavelength bounds that define regions in which absorption is formally
detected (``kinematic subsystems'') and calculates the rest-frame
equivalent width, $W_r(2796)$. The code also defines the absorption
redshift, $z_{\rm abs}$, as the median velocity of the apparent
optical depth distribution of {\MgII} absorption
\citep{cwc-thesis}. In Figures~\ref{fig:specEW}(a) and (b), we present
an example spectrum of quasar Q0235+164 with three {\MgII} kinematic
subsystems at $z_{\rm abs}=0.852$. The black histogram is the spectrum
for the {\MgII}~$\lambda 2796$ line (panel (a)) and for the
{\MgII}~$\lambda 2803$ line (panel (b)). The shaded regions in panel
(a) designate two of the three kinematic subsystems for this
system. For our TPCF analysis, we use only the pixels inside these
shaded regions.

We then fit all {\MgII} systems using Voigt profile (VP) decomposition
with {\sc Minfit} \citep{cwc-thesis, cv01, cvc03} and adopt the model
with the fewest statistically significant VP components. {\sc Minfit}
defines the VP component (cloud) velocities, column densities, and
Doppler $b$ parameters. Full details of {\sc Minfit} and the fitting
process are described in \citet{evans-thesis} and most VP fits are
presented in \citet{kcems11}. An example VP fit is presented in
Figures~\ref{fig:specEW}(a) and (b) as the thick red line. Individual
VP components are plotted as thin red lines centered at velocities
indicated by the red, vertical ticks. This system was fitted with
three components in the first shaded region, six in the second shaded
region, and one component at larger velocities.

The absorber data are listed in columns (9)--(12) of
Table~\ref{tab:props}\footnote{We have data for two additional
  {\magiicat} absorber--galaxy pairs but do not include them in the
  table nor the analysis. The first is an outlier with
  $W_r(2796)=4.422$~{\AA} and has no galaxy $B-K$. The second has
  $W_r(2796)=0.032$~{\AA}, which is lower than our detection
  threshold.}. The total column densities, $\log N({\MgII})$, in
column (11) are calculated by summing the column densities of each
cloud. For a few absorbers, at least one cloud in the absorber does
not have a well constrained column density. In these cases, we report
only the approximate column densities. Column (12) lists the reference
for the absorption data. In several instances, we fit the absorbers
for this work.

To account for differences in the quality of our spectra and to ensure
we can uniformly detect absorption throughout our sample, we use an
equivalent width detection threshold. We calculate the mean
$3\sigma~W_r(2796)$ detection threshold in each spectrum, defined as
the minimum $W_r(2796)$ a kinematic subsystem should have in order to
be detected. Figure~\ref{fig:specEW}(c) presents the cumulative
distribution function of the detection threshold in each spectrum in
our sample, which is $\sim95\%$ complete to roughly 0.04~{\AA} within
$\pm800$~{\kms} for all absorbers in our sample. We adopt this value
as our equivalent width detection threshold and do not include any
kinematic subsystems with $W_r(2796)<0.04$~{\AA} in our analysis. An
example kinematic subsystem that is just below our sensitivity limit
with $W_r(2796)=0.03$~{\AA} is presented in Figure~\ref{fig:specEW}(a)
at $v\sim 550$~{\kms}.

We also have an additional 23 absorber--galaxy pairs with HIRES/Keck
or UVES/VLT quasar spectra in which only an upper limit on the {\MgII}
equivalent width was measured, though we do not use the data in this
work as we cannot obtain kinematics information for these
``nonabsorbers.'' In all but one case, the upper limits on absorption
are lower than our adopted equivalent width detection
threshold. Therefore, if we were able to obtain kinematic information
for these systems, they would not be included in our sample as their
equivalent widths are too low.

\section{Characterizing Kinematics}
\label{sec:kinematics}

Several methods to examine the kinematics of {\MgII} absorbers have
been used in the literature; two common methods utilize the velocity
distribution and/or clustering of VP components. These include the
distributions of Voigt profile components and the two-point
correlation function.

\subsection{Voigt Profile Component Distributions}

\begin{figure}[ht]
\includegraphics[width=\linewidth]{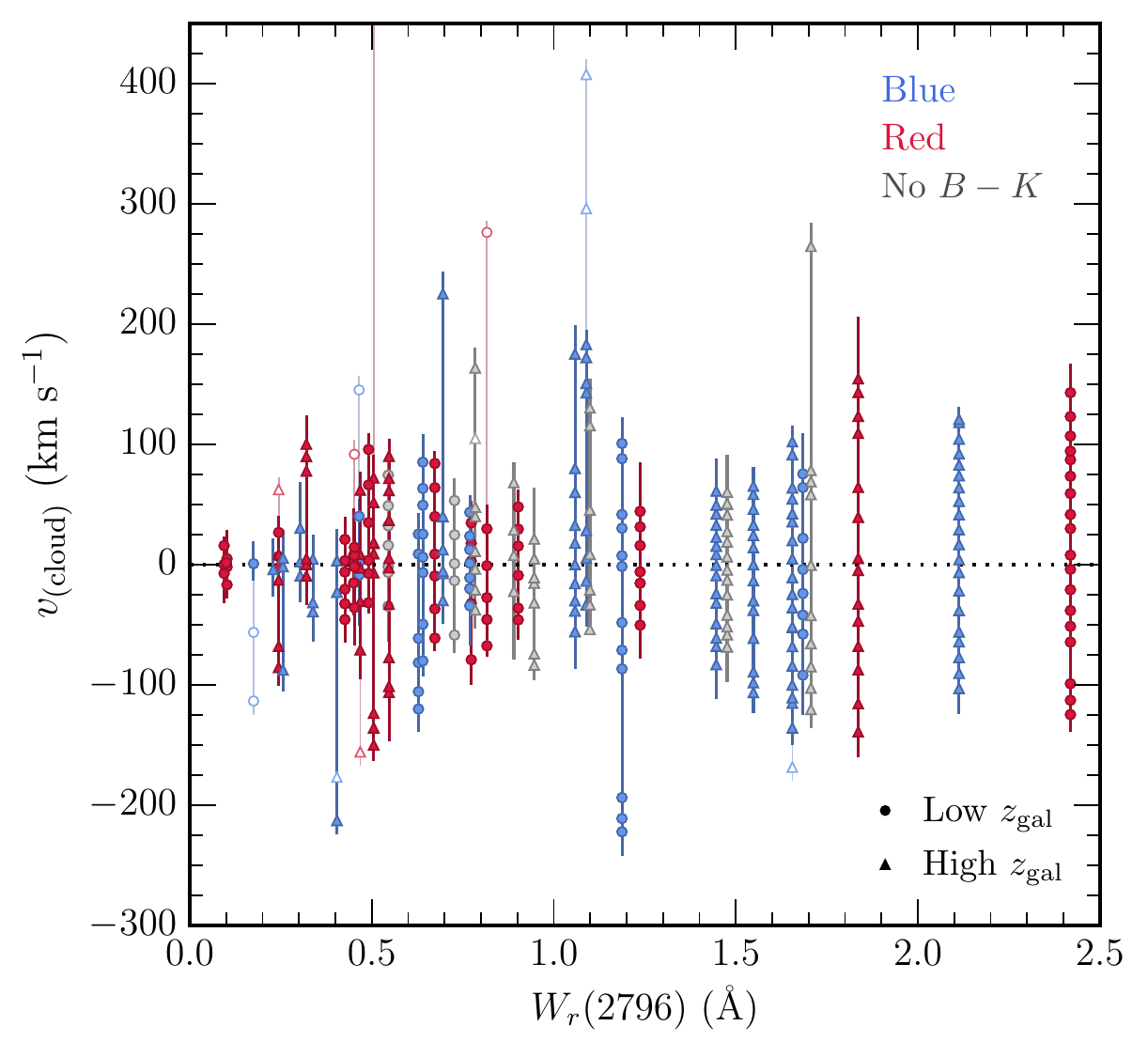}
\caption[]{Absorber velocity distribution as a function of rest
  equivalent width, $W_r(2796)$. Points represent each individual
  fitted cloud while vertical lines show the extremes in velocity of
  each profile, neglecting gaps between kinematic subsystems. Open
  points and lightly colored lines represent those subsystems which
  have been dropped from the sample due to our equivalent width
  detection sensitivity cut ($W_r(2796)\geq0.04$~{\AA}). See the top
  of Figure~\ref{fig:specEW}(a) for an example to translate points and
  vertical lines to a spectrum. Point colors and types represent our
  $B-K$ and $z_{\rm gal}$ subsamples, respectively. The point going
  off the panel at $W_r(2796)\sim0.5$~{\AA} is a cloud located at
  $v_{\rm cloud}\sim550$~{\kms} and belongs to the absorber presented
  in Figure~\ref{fig:specEW}. Large velocity spreads can be found for
  nearly the full $W_r(2796)$ range.}
\label{fig:vvsEW}
\end{figure}

\begin{figure*}[ht]
\centering
\includegraphics[width=\linewidth]{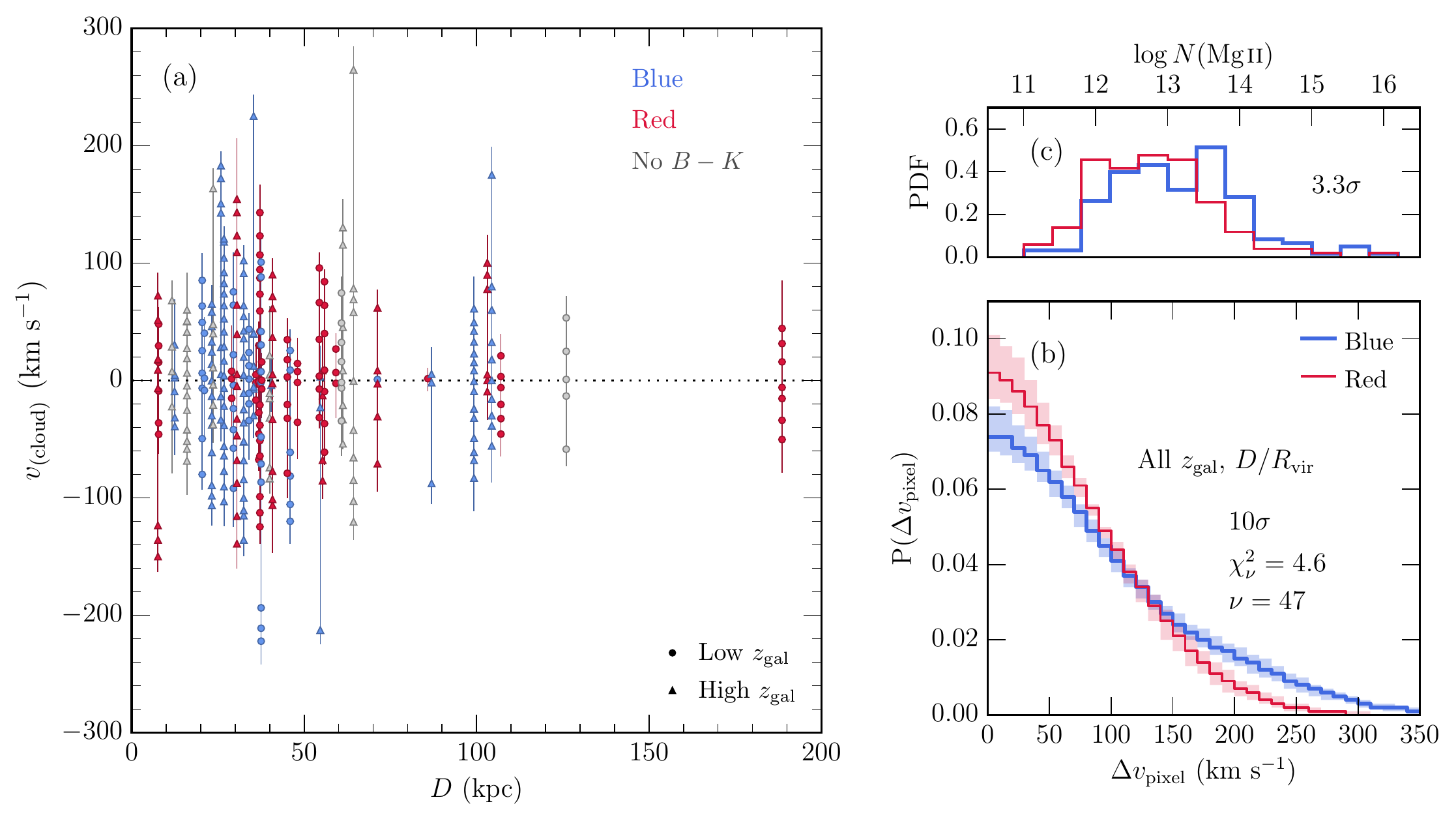}
\caption[]{(a) Velocity distribution of clouds (Voigt profile
  components) in the absorption profiles as a function of impact
  parameter. Points and lines are similar to
  Figure~\ref{fig:vvsEW}. Most absorption is found within $|v_{\rm
    (cloud)}|=150$~{\kms} of $z_{\rm abs}$ (which defines the velocity
  zero point from the optical depth weighted median of absorption) and
  absorbers located at smaller impact parameters appear to have
  broader profiles in velocity space than those further away. (b)
  Pixel-velocity two-point correlation functions for blue and red
  galaxies (solid lines). Shaded regions represent the $1\sigma$
  uncertainties around the data from a bootstrap analysis. The
  significance of a chi-squared test comparing the distributions of
  the blue and red TPCFs is listed in the panel. We find that blue
  galaxies have larger velocity dispersions than red galaxies. (c)
  Cloud column density distributions for blue and red galaxies. The
  counts in each bin have been normalized by the total number of
  clouds for the subsample. The results of a KS test comparing the
  plotted subsamples is listed in the panel. Blue galaxies tend to
  have larger column densities than red galaxies at the $3.3\sigma$
  level.}
\label{fig:bivariate}
\end{figure*}

In Figure~\ref{fig:vvsEW}, we present the kinematics of our {\MgII}
absorbers as a function of rest equivalent width, $W_r(2796)$, for all
39 absorber--galaxy pairs, including an additional eight pairs for
which we have a high-resolution quasar spectrum but no measured $B-K$
value. We show a simplified spectrum of each {\MgII} absorber with a
velocity zero point at $z_{\rm abs}$, defined as the optical depth
weighted median of the absorption. Clouds (VP components) are plotted
as points centered at their fitted line-of-sight velocity (see
Figure~\ref{fig:specEW}(a)) and total $W_r(2796)$ of the associated
absorber. The spread in velocity of the absorbers is plotted as
vertical lines and represents only the largest deviations from $z_{\rm
  abs}$; gaps between kinematic subsystems (i.e., stretches of
continuum within the extreme velocity bounds of the absorber) are not
presented here. Kinematic subsystems whose equivalent widths are below
our sensitivity cut are plotted as open points with lighter vertical
lines. Point colors indicate the rest-frame $B-K$ color of the host
galaxy (a proxy for star formation rate), with blue points
representing galaxies with {\blue}, red points as galaxies with
{\red}, and gray points are those galaxies for which we do not have a
$B-K$ measurement (8 galaxies). Point types indicate whether the host
galaxy is located at low redshift (circles, {\lowz}), or high redshift
(triangles, {\highz}).

As shown in Figure~\ref{fig:vvsEW}, large absorber velocity spreads
can be found for absorbers of nearly all equivalent width strengths,
especially when an equivalent width detection threshold is not
enforced (open points). The narrowing of the profiles near
$W_r(2796)\sim 1.0-1.2$~{\AA} is due to the absorbers becoming
saturated, which occurs for $\log N({\MgII})\sim 13$. Below this
point, large velocity spreads are largely due to several kinematic
subsystems spread out in velocity and may have stretches of continuum
between subsystems. This is most obvious in the absorber presented in
Figure~\ref{fig:specEW}(a), which, when including kinematic subsystems
below our detection threshold, has the largest velocity spread in the
sample. Above this point the number of clouds fit to the profile
increases and corresponds to an increasing velocity width. The
degeneracy between velocity spread and equivalent width in this plot
shows that equivalent width is a poor indicator of absorber
kinematics.

We present the kinematics of our {\MgII} absorbers as a function of
galaxy rest-frame $B-K$ color, redshift, $z_{\rm gal}$, and impact
parameter, $D$, in Figure~\ref{fig:bivariate}(a). Point colors and
types, along with line colors are similar to those in
Figure~\ref{fig:vvsEW}. We find several qualitative trends in these
results. Clouds are mostly found within $|v_{\rm (cloud)}| =
150$~{\kms} of the absorber systemic velocity. As absorption is probed
further from the galaxy (moving outward with increasing $D$), the
velocity spread of absorption may decrease from large velocity spreads
at low $D$ to smaller velocity spreads at higher $D$. The absorbers
may be more extended in velocity for blue galaxies than red. Also, it
appears that most of the highest velocity clouds are located around
galaxies at high redshift.

This method of examining the absorption kinematics has been used
often, though with the velocities shifted to the galaxy systemic
velocity \citep[see e.g.,][]{tumlinson13, werk13, mathes14}. Though
the method is effective, it is difficult to extract clear kinematic
trends with, for example, galaxy redshift and color, let alone
characterizing the kinematics of the gas itself (rather than with
respect to the galaxy).

\subsection{Pixel-velocity Two-point Correlation Functions}
\label{sec:TPCF}

\begin{deluxetable*}{lccccrlllll}
\tablecolumns{11} 
\tablewidth{0pt} 
\tablecaption{TPCF {\vfifty} and {\vninety}
  Measurements \label{tab:v50}}
\tablehead{
  \colhead{} &
  \colhead{} &
  \colhead{} &
  \colhead{} &
  \colhead{} &
  \colhead{} &
  \multicolumn{2}{c}{$\Delta v_{\rm pixel}$} &
  \colhead{} &
  \multicolumn{2}{c}{$\Delta(v_{\rm pixel}/V_{\rm circ})$} \\
  \cline{7-8} \cline{10-11} \\[-5pt]
  \colhead{Sample} &
  \colhead{Cut} &
  \colhead{Cut} &
  \colhead{$\langle z_{\rm gal} \rangle$} &
  \colhead{$\langle D/R_{\rm vir} \rangle$} &
  \colhead{\# Gals} &
  \colhead{{\vfifty}\tablenotemark{a}}  &
  \colhead{{\vninety}\tablenotemark{a}} &
  \colhead{}                            &
  \colhead{{\vfifty}}  &
  \colhead{{\vninety}} 
}
\startdata
\cutinhead{Figure~\ref{fig:bivariate}(b) and Figure~\ref{fig:bivariatenorm}}\\[-3pt]
Blue & {\blue} & $\cdots$ & 0.754 & 0.26 & 19 & \phn$75_{-6}^{+5}$ & $205_{-17}^{+12}$ && $0.63_{-0.07}^{+0.06}$ & $1.76_{-0.18}^{+0.15}$ \\[3pt]
Red  & {\red}  & $\cdots$ & 0.537 & 0.25 & 20 & \phn$60_{-6}^{+5}$ & $151_{-14}^{+11}$ && $0.35_{-0.04}^{+0.03}$ & $0.87_{-0.09}^{+0.07}$ \\[3pt]

\cutinhead{Figure~\ref{fig:BKzabs} and Figure~\ref{fig:BKzabsnorm}}\\[-3pt]
Blue -- Low $z_{\rm gal}$  & {\blue} & {\lowz}  & 0.472 & 0.27 &  7 & \phn$78_{-13}^{+9}$ & $199_{-35}^{+21}$ && $0.70_{-0.14}^{+0.10}$ & $1.84_{-0.36}^{+0.23}$ \\[3pt]
Blue -- High $z_{\rm gal}$ & {\blue} & {\highz} & 0.783 & 0.25 & 12 & \phn$71_{-6}^{+4}$  & $200_{-15}^{+11}$ && $0.58_{-0.07}^{+0.05}$ & $1.61_{-0.16}^{+0.14}$ \\[3pt]
Red -- Low $z_{\rm gal}$   & {\red}  & {\lowz}  & 0.472 & 0.27 & 14 & \phn$49_{-8}^{+7}$  & $123_{-23}^{+16}$ && $0.29_{-0.05}^{+0.04}$ & $0.73_{-0.11}^{+0.07}$ \\[3pt]
Red -- High $z_{\rm gal}$  & {\red}  & {\highz} & 0.767 & 0.17 &  6 & \phn$73_{-7}^{+6}$  & $177_{-15}^{+13}$ && $0.41_{-0.05}^{+0.04}$ & $1.00_{-0.12}^{+0.10}$ \\[3pt]

\cutinhead{Figure~\ref{fig:BKDRabs} and Figure~\ref{fig:BKDRabsnorm}}\\[-3pt]
Blue -- Low $D/R_{\rm vir}$  & {\blue} & {\lowdr}  & 0.780 & 0.13 &  9 & \phn$ 68_{- 8}^{+ 5}$ & $193_{- 19}^{+ 14}$ && $0.58_{-0.06}^{+0.04}$ & $1.59_{-0.16}^{+0.15}$ \\[3pt]
Blue -- High $D/R_{\rm vir}$ & {\blue} & {\highdr} & 0.656 & 0.28 & 10 & \phn$ 77_{- 9}^{+ 6}$ & $199_{- 22}^{+ 15}$ && $0.67_{-0.11}^{+0.09}$ & $1.76_{-0.28}^{+0.20}$ \\[3pt]
Red -- Low $D/R_{\rm vir}$   & {\red}  & {\lowdr}  & 0.637 & 0.14 & 10 & \phn$ 72_{- 9}^{+ 7}$ & $174_{- 16}^{+ 14}$ && $0.37_{-0.07}^{+0.05}$ & $0.95_{-0.14}^{+0.11}$ \\[3pt]
Red -- High $D/R_{\rm vir}$  & {\red}  & {\highdr} & 0.459 & 0.31 & 10 & \phn$ 48_{- 3}^{+ 2}$ & $114_{-  7}^{+  4}$ && $0.31_{-0.03}^{+0.02}$ & $0.75_{-0.06}^{+0.04}$ \\[-5pt]
\enddata 
\tablenotetext{a}{{\kms}} 
\end{deluxetable*}

We extend beyond the line of work with cloud velocities started by
\citet{pb90} by examining the pixel-velocity two-point correlation
function (TPCF) for various galaxy subsamples and compare the
resultant line-of-sight velocity dispersions. Previous works
constructed TPCFs using cloud (VP component; the ticks at the top of
Figure~\ref{fig:specEW}(a)) velocities, while we use the velocities of
pixels in regions of the spectrum which contribute to the overall
{\MgII} equivalent width (i.e., kinematic subsystems, see shaded
regions in Figure~\ref{fig:specEW} for examples). Compared to cloud
velocities, pixel velocities better represent the spread in
absorption, provide more velocity pairs for better statistics, and can
be compared more easily to simulations because the absorption profiles
do not need to be Voigt profile modeled. They are also
model-independent, i.e., they do not depend on the fitting method used
and the resulting fit. We study the velocity dispersions of the
absorbers for various galaxy subsamples using the pixel-velocity TPCF.

To first construct the TPCF, which is a measure of the internal
absorber velocity dispersion, we define a subsample of galaxies. From
the spectra of background quasars associated with these galaxies, we
obtain the velocities (where $v=0$~{\kms} represents $z_{\rm abs}$,
the optical depth weighted median of absorption) of pixels located in
regions in which {\MgII} absorption has been formally detected, also
known as kinematic subsystems (gray shaded regions of
Figure~\ref{fig:specEW}(a)). The velocity bounds of these regions are
defined by searching the spectrum redwards and bluewards from the
subsystem velocity centroid until the significance in the per pixel
equivalent width drops below $1\sigma$ \citep{cv01}. 

We then combine the pixel velocities from each absorber--galaxy pair
in the subsample. This method uses the experiment in which combining a
single absorbing sightline around multiple galaxies for a certain
subsample is equivalent to multiple absorbing sightlines around a
single galaxy of the same type as the subsample. This allows for more
robust statistics and prevents any possible bias that might arise from
the absorbers with the most pixels. Then the absolute value of the
velocity separations is calculated between each possible pixel
velocity pair for the subsample to get $\Delta v_{\rm pixel}$. The
TPCF is created by binning the velocity separations and normalizing
the value in each bin by the total number of pixel velocity pairs in
the subsample to account for differing numbers of pixels in each
subsample when comparing between subsamples. The TPCF is therefore a
probability distribution function. We use a bin size of 10~{\kms},
which corresponds to roughly one resolution element of both the
HIRES/Keck and UVES/VLT spectrographs (three pixels per resolution
element, with a FWHM resolution of $\sim 6.6$~{\kms}).

To determine the uncertainties on the TPCFs, we conduct a bootstrap
analysis. We randomly draw with replacement a sample of kinematic
subsystems from the subsample we are examining which contains the same
number of kinematic subsystems as the original data and construct a
TPCF. We run 1000 bootstrap realizations and then calculate the
$1\sigma$ standard deviations from the mean of the realizations in
each TPCF bin. The bootstrap uncertainties are plotted as shaded
regions around the TPCF. 

We also characterize the TPCFs by measuring the velocity separations
within which 50\% and 90\% of the data reside, {\vfifty} and
{\vninety}, respectively. For these TPCFs, $v$ in {\vfifty} and
{\vninety} represents $v_{\rm pixel}$. Uncertainties on these values
are obtained from the bootstrap analysis and represent $1\sigma$
deviations from the mean. These values and their uncertainties are
presented in Table~\ref{tab:v50} for each subsample.

We present the absorber TPCFs comparing blue and red galaxies in
Figure~\ref{fig:bivariate}(b). Blue galaxies are presented as the
thick blue line with blue shaded areas indicating the bootstrap
uncertainties, while the thin red line and shading represents red
galaxies. In this panel, we find that the absorption associated with
blue galaxies has a larger velocity dispersion than with red galaxies,
which can also be seen in panel (a). To test whether the two samples
were drawn from the same population, we ran a chi-squared test on the
binned TPCFs (including the uncertainties on the TPCF) and find that
the null hypothesis that the samples were drawn from the same
population can be ruled out at the $10\sigma$ level. The significance,
$\sigma$, the reduced chi-squared, $\chi^2_{\nu}$, and the number of
degrees of freedom, $\nu$, is presented in
Figure~\ref{fig:bivariate}(b). We find that both the {\vfifty} and
{\vninety} measurements for blue galaxies (75~\kms and 205~\kms,
respectively) are greater than for red galaxies (60~\kms and 151~\kms,
respectively). This indicates that absorbers around blue galaxies have
significantly larger velocity dispersions than those around red
galaxies.

\subsection{Cloud Column Densities}

\begin{figure}[ht]
\centering
\includegraphics[width=\linewidth]{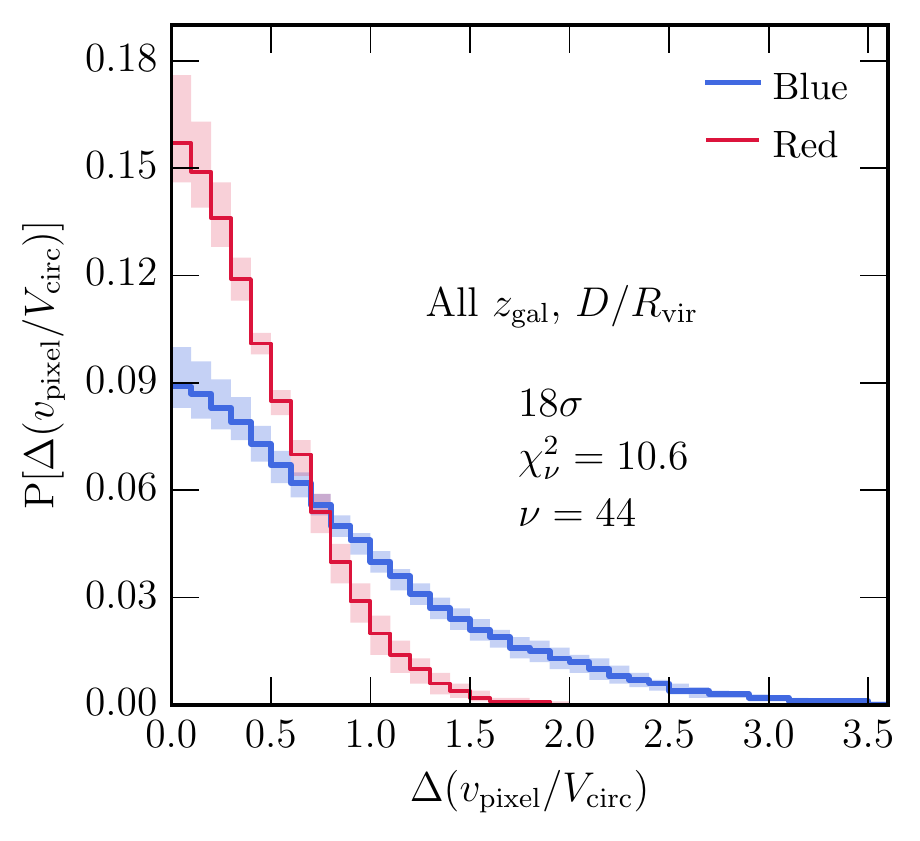}
\caption[]{Pixel-velocity TPCFs normalized by the host galaxy circular
  velocity (to account for galaxy halo mass) for blue and red
  galaxies. Normalizing the velocities by $V_{\rm circ}$ does not
  change the general result that absorbers around blue galaxies have
  larger velocity dispersions than around red galaxies, which was also
  presented in Figure~\ref{fig:bivariate}(b). However, the result
  becomes more significant at the $18\sigma$ level.}
\label{fig:bivariatenorm}
\end{figure}

Examining the cloud column densities in addition to the absorption
velocity dispersions yields a more complete picture of the physics
involved in placing and maintaining {\MgII} absorption in the halos of
galaxies. This is especially true considering that column densities
depend on the ionization conditions, temperature, metallicity, and
path length of the gas that is being probed.  As stated in
Section~\ref{sec:spectra}, we obtain cloud column densities for
absorbers using Voigt profile decomposition. We examine the column
density distributions for the same subsamples we use for the TPCFs in
order to obtain a more complete picture of the gas properties as a
function of galaxy properties, and therefore, evolutionary processes.

In Figure~\ref{fig:bivariate}(c), we present the cloud column density
distributions for blue and red galaxy subsamples. The counts in each
column density bin are normalized to the total number of clouds in
each subsample to create a probability distribution function. We use
the Kolmogorov--Smirnov (KS) test comparing the plotted column density
distributions to determine if the two samples were drawn from the same
population. With a $3.3\sigma$ significance, we find that the null
hypothesis that the cloud column densities for blue and red galaxies
were drawn from the same population can be ruled out. Thus, absorption
associated with blue galaxies tends to have larger column densities
than absorption associated with red galaxies.

\subsection{Mass-normalized Pixel-velocity TPCF}

To account for the mass of the galaxy hosting absorption, we normalize
the pixel velocities by the maximum circular velocity, $V_{\rm circ}$,
of the host galaxy. We do this because our sample spans a range of
galaxy halo masses and because of the fact that our red galaxies tend
to be more massive than our blue galaxies (see
section~\ref{sec:methods} and Figure~\ref{fig:BKMh}). Here we present
methods for constructing normalized absorber TPCFs.

\begin{figure*}[ht]
\includegraphics[width=\linewidth]{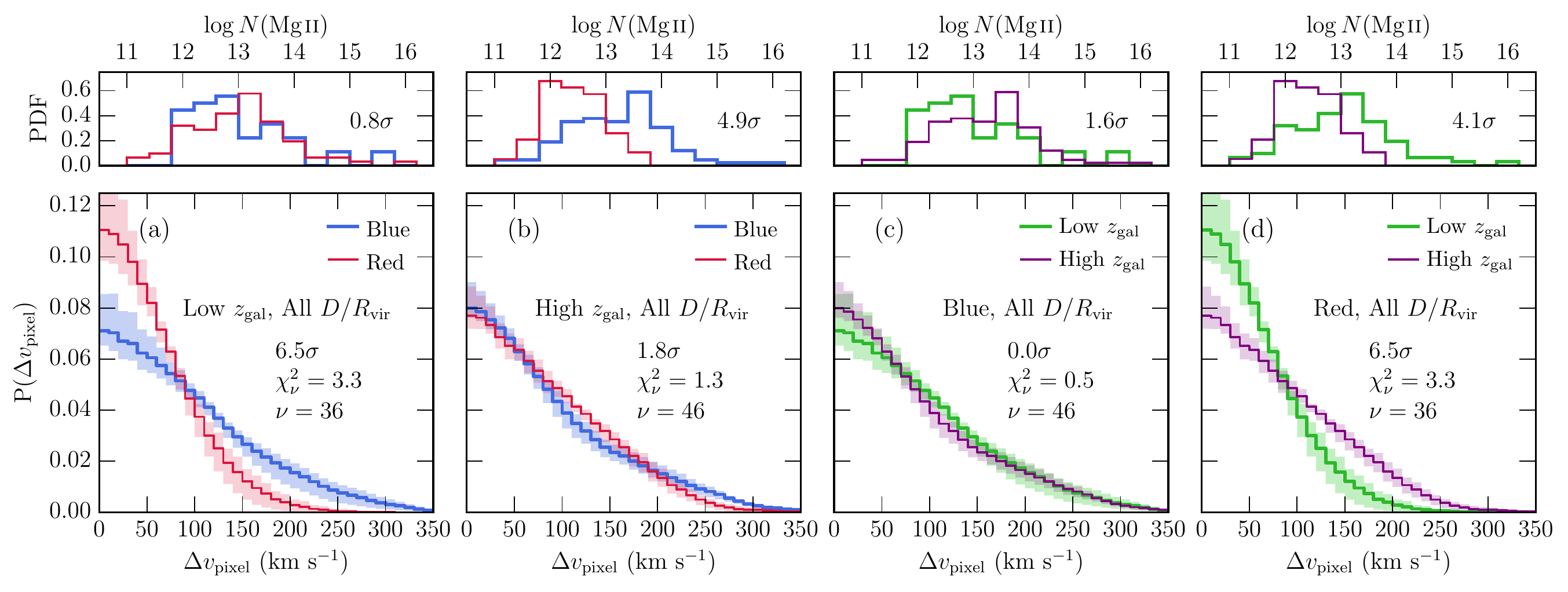}
\caption[]{Pixel-velocity two-point correlation functions for color
  and redshift subsamples in the bottom panels. Solid lines are the
  TPCFs while shaded regions are the $1\sigma$ bootstrap
  uncertainties. We list the significance, the reduced chi-squared
  value, $\chi^2_{\nu}$, and the degrees of freedom, $\nu$, from a
  chi-squared test comparing subsamples in each panel. In each case,
  blue galaxies have {\blue} and red galaxies have {\red}. We find
  that red, low $z_{\rm gal}$ ({\lowz}) galaxies have significantly
  smaller absorber velocity dispersions than red, high $z_{\rm gal}$
  ({\highz}) galaxies or blue galaxies (panels (d) and (a)), whereas
  there are no differences in the velocity dispersions for blue
  galaxies (panel (c)) or for blue and red galaxies at high $z_{\rm
    gal}$ (panel (b)). The panels above the TPCFs present the cloud
  column density distributions for the same TPCF subsamples. The
  plotted significance in each panel is the result of a KS test
  between subsamples. The column densities only differ with redshift
  for red galaxies, with smaller column densities at higher
  redshifts.}
\label{fig:BKzabs}
\end{figure*}

\begin{figure*}[ht]
\includegraphics[width=\linewidth]{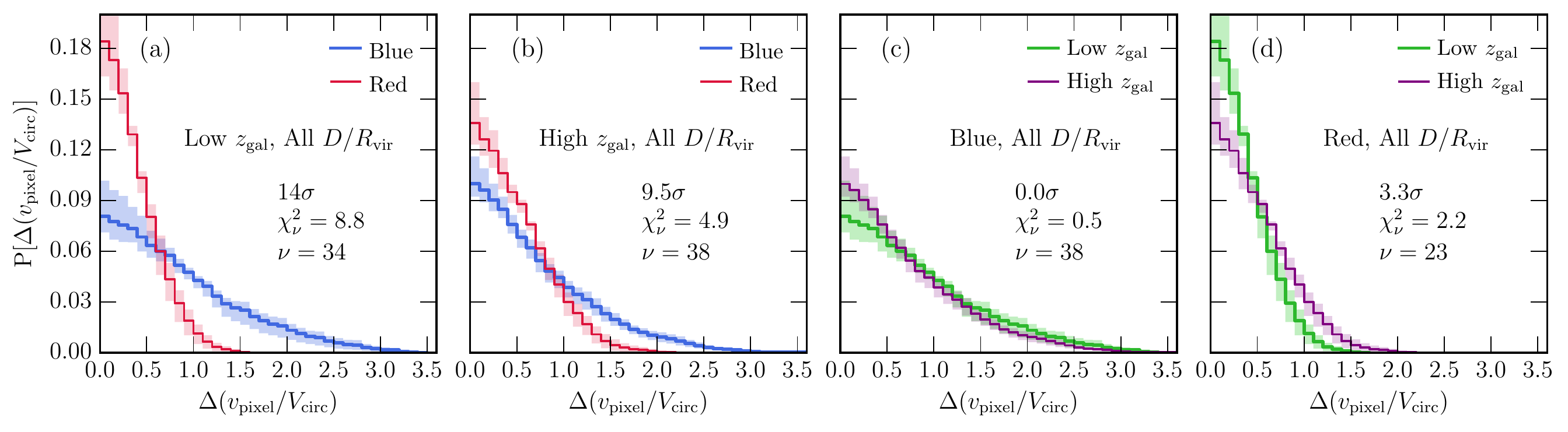}
\caption[]{Mass-normalized TPCFs for the same color and redshift
  subsamples as those in Figure~\ref{fig:BKzabs}. Normalizing the
  velocities by $V_{\rm circ}$ does not change the general results
  compared to those presented in Figure~\ref{fig:BKzabs}, with the
  exception of the high redshift samples in panel (b). In this panel,
  we find that blue galaxies have a larger absorber velocity
  dispersion than red galaxies, in contrast to no difference for the
  unnormalized TPCFs. The redshift evolution of red galaxies in panel
  (d) becomes less significant with a value of $3.3\sigma$.}
\label{fig:BKzabsnorm}
\end{figure*}

\begin{figure*}[ht]
\includegraphics[width=\linewidth]{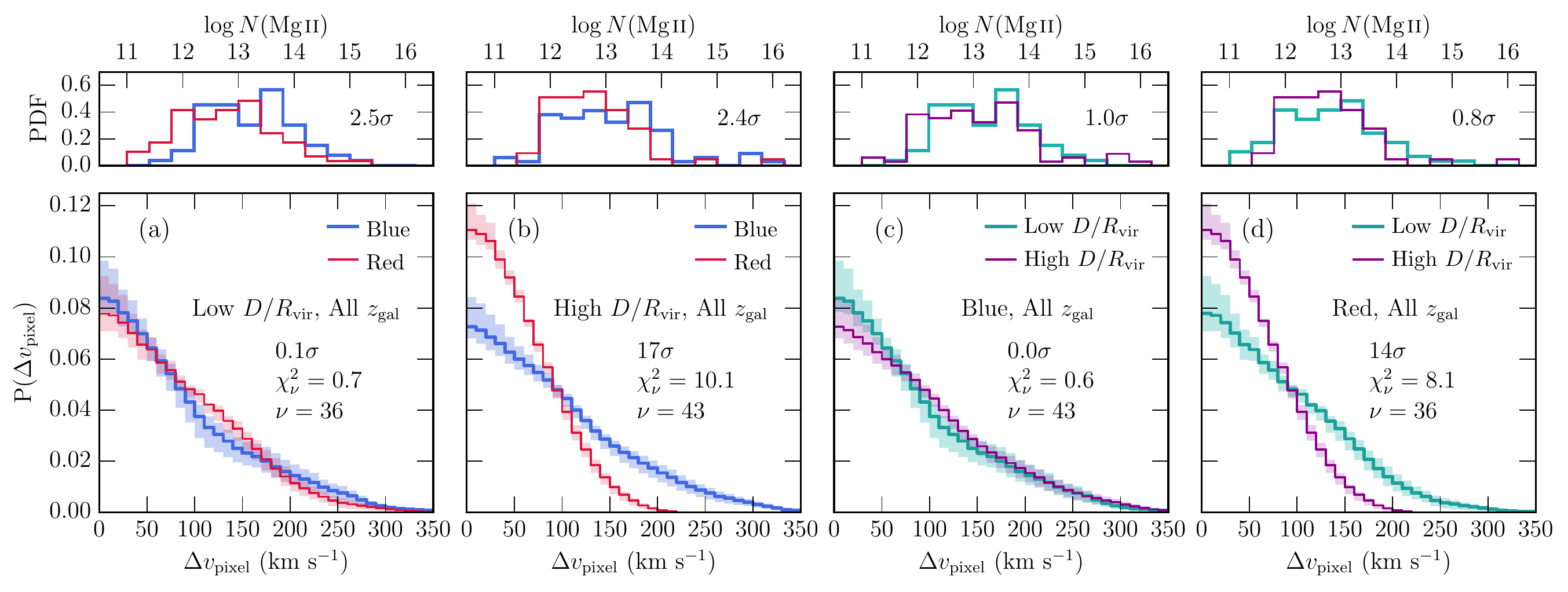}
\caption[]{Pixel-velocity two-point correlation functions for
  subsamples sliced by galaxy color, $B-K$, and virial
  radius-normalized impact parameter, $D/R_{\rm vir}$, in the lower
  panels. Lines, shading, and chi-squared test results are similar to
  those listed in Figure~\ref{fig:BKzabs}. We find that the red, low
  $D/R_{\rm vir}$ subsample is a significant outlier such that it has
  a smaller absorber velocity dispersion than either blue galaxies
  ($17\sigma$ in panel (b)) or red, high $D/R_{\rm vir}$ galaxies
  ($14\sigma$ in panel (d)). Upper panels display the differences
  between cloud column density distributions for the same subsamples
  plotted in the TPCF panels. KS test results between plotted
  distributions are noted in each panel. We find no dependence of the
  cloud column densities with color and $D/R_{\rm vir}$.}
\label{fig:BKDRabs}
\end{figure*}

\begin{figure*}[ht]
\includegraphics[width=\linewidth]{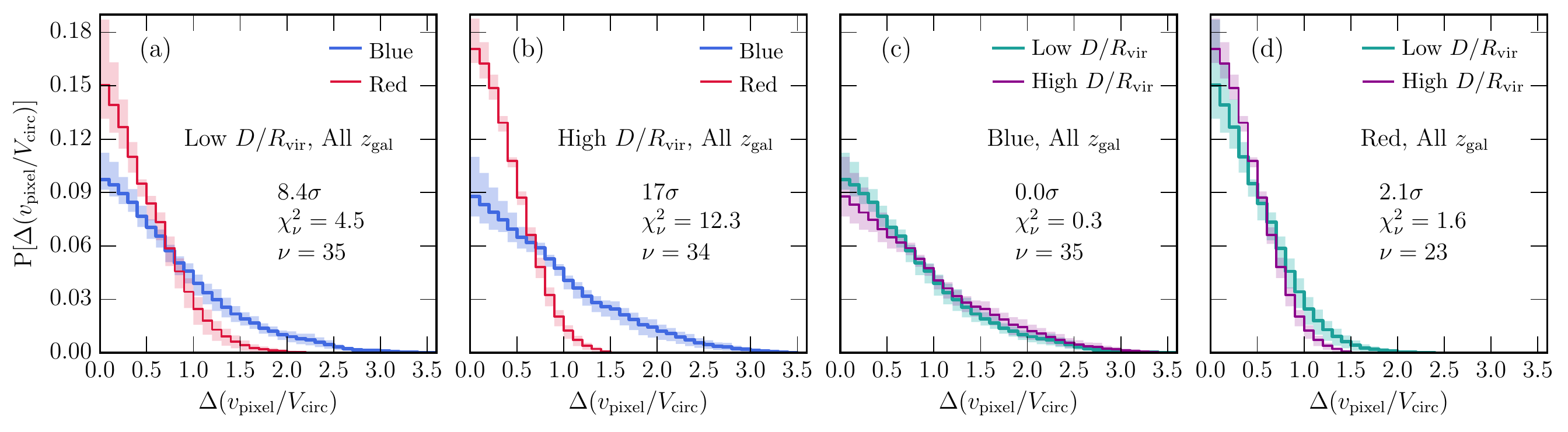}
\caption[]{Mass-normalized TPCFs for the subsamples plotted in
  Figure~\ref{fig:BKDRabs}. Contrary to the results with the
  unnormalized absorber TPCFs, we find no dependence on $D/R_{\rm
    vir}$ for either blue or red galaxies in panels (c) and (d),
  respectively, however red galaxies have smaller absorber velocity
  dispersions than blue galaxies at all $D/R_{\rm vir}$ (panels (a)
  and (b)).}
\label{fig:BKDRabsnorm}
\end{figure*}

We calculate the mass-normalized absorber TPCF using a similar
procedure as the unnormalized absorber TPCF. Before we calculate the
pixel pair velocity separations for a given subsample, we normalize
each pixel velocity by the $V_{\rm circ}$ of the host galaxy. After
determining the velocity separations, $\Delta (v_{\rm pixel}/V_{\rm
  circ})$, we bin the values using the same methods used for the
unnormalized TPCFs. The normalized absorber TPCF for blue and red
galaxies is presented in Figure~\ref{fig:bivariatenorm}. The general
result in this panel that blue galaxies have a larger absorber
velocity dispersion than red galaxies does not differ from the
unnormalized absorber TPCF in Figure~\ref{fig:bivariate}(b), but the
significance of the chi-squared test is greater here. We also present
measurements of {\vfifty} and {\vninety} for each subsample in the
right-most columns of Table~\ref{tab:v50}, where the $v$ in this case
represents $(v_{\rm pixel}/V_{\rm circ})$. For these subsamples, we
find that the values of {\vfifty} and {\vninety} are very different,
with much larger values for the blue subsample (0.6 and 1.8,
respectively) than the red subsample (0.35 and 0.9, respectively); the
values for the blue subsample are roughly twice as large as for the
red subsample.

\section{Multivariate Analysis}
\label{sec:results}

In this section, we report on a multivariate analysis of the
kinematics and column density distributions for blue and red galaxies
cut by (1) galaxy redshift, $z_{\rm gal}$, and (2) the projected
radial distance normalized by the virial radius, $D/R_{\rm vir}$.

\subsection{Redshift Evolution}
\label{sec:redshift}

While we find significant differences between blue and red galaxies in
the TPCFs in Figure~\ref{fig:bivariate}(b), the differences may be
washed out by other effects. One such effect is the fact that the star
formation rate has decreased over time to the present day from a peak
at $z \sim 2-3$ \citep[e.g.,][]{hopkins06}. Therefore, we slice our
blue and red subsamples into low and high $z_{\rm gal}$, using a
median cut of $\langle z_{\rm gal} \rangle = 0.656$. The mean redshift
of the low $z_{\rm gal}$ subsample is $z_{\rm gal}=0.469$, while it is
$z_{\rm gal}=0.804$ for the high $z_{\rm gal}$ subsample,
corresponding roughly to a 2~Gyrs time span between mean redshifts. We
present the TPCFs for $B-K$ and $z_{\rm gal}$ subsamples in
Figure~\ref{fig:BKzabs}. In the panels, we list the significance of a
chi-squared test between subsample pairs as well as the reduced
chi-squared value, $\chi^2_{\nu}$, and the degrees of freedom, $\nu$
for each panel. Measurements of {\vfifty} and {\vninety} for each
subsample are presented in Table~\ref{tab:v50}.

We find that the TPCF for the red, low $z_{\rm gal}$ subsample is an
outlier such that it has a significantly smaller velocity dispersion
than the rest of the subsamples. In this case, the absorber velocity
dispersion for red galaxies evolves with redshift over a span of
roughly 2~Gyrs ($6.5\sigma$, Figure~\ref{fig:BKzabs}(d)), while there
is no such evolution for blue galaxies ($0.0\sigma$,
Figure~\ref{fig:BKzabs}(c)). At high $z_{\rm gal}$ in
Figure~\ref{fig:BKzabs}(b), absorption in red galaxies has similar
velocity dispersions as in blue galaxies ($1.8\sigma$). However, at
lower $z_{\rm gal}$ in Figure~\ref{fig:BKzabs}(a), the velocity
dispersion for red galaxies decreases, whereas the dispersion for blue
galaxies remains constant ($6.5\sigma$). The values of {\vfifty} and
{\vninety} for all TPCFs are consistent within uncertainties ($\sim
75$~\kms and $\sim 190$~\kms, respectively) except the red, low
$z_{\rm gal}$ subsample, which has smaller values than the rest of the
subsamples ($\sim 50$~\kms and $\sim 120$~\kms, respectively) and thus
confirms the low velocity dispersions.

In the panels above the TPCFs we plot the cloud column density
distributions for the same subsamples as those in the TPCF plots. The
listed significance is the result of a KS test between plotted
subsamples. The outlying subsample in these panels is the red, high
$z_{\rm gal}$ subsample which has smaller values of $\log N({\MgII})$
than both the blue, high $z_{\rm gal}$ subsample ($4.9\sigma$, panel
(b)), and the red, low $z_{\rm gal}$ subsample ($4.1\sigma$, panel
(d)). We find no difference between the column density distributions
associated with blue and red galaxies at low $z_{\rm gal}$
($0.8\sigma$, panel (a)) nor do we find redshift evolution in the
column density distributions for blue galaxies ($1.6\sigma$, panel
(c)).

Figure~\ref{fig:BKzabsnorm} presents TPCFs in which the pixel
velocities have been normalized by the circular velocity of the host
galaxy. Plotted subsamples are the same as those in
Figure~\ref{fig:BKzabs}. We do not plot the column density
distributions above these TPCF panels because the act of normalizing
the velocities by $V_{\rm circ}$ does not affect the column
densities. In general, we find the same results for the normalized
TPCFs as we did in the unnormalized TPCFs. However, we find that the
redshift evolution present in the red galaxies (panel (d)) is no
longer as strong as it was when the pixel velocities were not
normalized ($3.3\sigma$). For both low and high $z_{\rm gal}$, red
galaxies tend to host absorbers with lower velocity dispersions than
blue galaxies at the $14\sigma$ (panel (a)) and $9.5\sigma$ (panel
(b)) levels, respectively. Lastly, we find that the velocity
dispersion of absorbers hosted by blue galaxies does not evolve with
redshift with a $0\sigma$ significance in panel (c).

{\it To summarize}, we find redshift evolution in both the velocity
dispersions and cloud column densities for absorbers associated with
red galaxies. However, the sense of the evolution is reversed in that
the velocity dispersion decreases from higher to lower $z_{\rm gal}$,
while the column densities increase for the same time span. We find no
evolution in either the velocity dispersion or the cloud column
densities for absorbers associated with blue galaxies.

\subsection{Radial Dependence}
\label{sec:DR}

Another effect that may be washing out differences in the TPCFs of
blue and red galaxies is the projected radial distance at which
absorption is found. Many previous works have studied the well-known
anti-correlation between $W_r(2796)$ and $D$, which is significant to
the $7.9\sigma$ level \citep[see][and references
  therein]{magiicat1}. Furthermore, since galaxies span a range of
masses, \citet{cwc-masses} normalized $D$ by the virial radius to
account for the mass of the host galaxy and found an even stronger
anti-correlation between $W_r(2796)$ and $D/R_{\rm vir}$ ($8.9\sigma$)
\citep[also see][]{magiicat3}. Since $W_r(2796)$ depends on column
densities and/or velocity spreads, examining the TPCFs and cloud
column densities as a function of $D/R_{\rm vir}$ may provide insight
into what aspect of the gas physics gives rise to the $W_r(2796)$ and
$D/R_{\rm vir}$ anti-correlation. Therefore, we present TPCFs for
subsamples sliced by median values of $\langle B-K \rangle = 1.4$ and
$\langle D/R_{\rm vir} \rangle = 0.24$ in
Figure~\ref{fig:BKDRabs}. The corresponding {\vfifty} and {\vninety}
measurements are listed in Table~\ref{tab:v50}.

We find that the internal velocity dispersion of absorbers (TPCF)
around blue galaxies does not depend on where the absorbers are
located in projected distance away from the galaxy ($0\sigma$, panel
(c)), except the dispersion does if the absorbers are located around
red galaxies ($14\sigma$, panel (d)). In red galaxies, the internal
dispersion of absorbers at low $D/R_{\rm vir}$ is comparable to
absorbers in blue galaxies, regardless of where they are being probed
($0.1\sigma$, panel (a)). The outlier of these TPCFs is the high
$D/R_{\rm vir}$, red galaxy subsample, which has a significantly
smaller velocity dispersion than for blue, high $D/R_{\rm vir}$
galaxies ($17\sigma$, panel (b)), or for red, low $D/R_{\rm vir}$
galaxies ($14\sigma$, panel (d)). These results are also represented
in the {\vfifty} and {\vninety} measurements, where all subsamples but
the red, high $D/R_{\rm vir}$ subsample have values of {\vfifty} and
{\vninety} that are consistent within uncertainties ($\sim 70$~\kms
and $\sim 190$~\kms, respectively). The red, high $D/R_{\rm vir}$
subsample has values that are lower than the rest of the subsamples
({\vfifty}$\sim 50$~\kms and {\vninety}$\sim 110$~\kms). We note that,
although these results are similar to those examining redshift
evolution in which one subsample is a clear outlier from the rest, we
find no significant anti-correlation from a Kendall-$\tau$ rank
correlation test between $z_{\rm gal}$ and $D/R_{\rm vir}$
($2.0\sigma$).

The column density distributions for the TPCF subsamples are plotted
above each panel in Figure~\ref{fig:BKDRabs}. Unlike the galaxy color
and redshift subsamples, we find no differences in the column
densities with color or $D/R_{\rm vir}$. The largest significance from
a KS test is between blue and red galaxies at low $D/R_{\rm vir}$ with
$2.5\sigma$, where red galaxies may tend to have smaller cloud column
distributions than blue galaxies. This trend may also be present at
high $D/R_{\rm vir}$ with a $2.4\sigma$ significance. Comparing low
and high $D/R_{\rm vir}$ for blue galaxies and red galaxies, we find
insignificant results from the KS test, with $1.0\sigma$ and
$0.8\sigma$, respectively.

We present the TPCFs normalized by the host galaxy virial radius in
Figure~\ref{fig:BKDRabsnorm} for the same $B-K$ and $D/R_{\rm vir}$
subsamples as in Figure~\ref{fig:BKDRabs}. Here we find that red
galaxies have lower velocity dispersions than blue galaxies at all
$D/R_{\rm vir}$, with a significance level of $8.4\sigma$ at low
$D/R_{\rm vir}$ in panel (a) and $17\sigma$ at high $D/R_{\rm vir}$ in
panel (b). The {\vfifty} and {\vninety} values for these subsample
pairs are also not consistent within uncertainties, where the
{\vfifty} values for blue galaxies ($\sim 0.6$) are roughly twice as
large as those for red galaxies ($\sim 0.3$) for all $D/R_{\rm
  vir}$. In panel (c) we find no difference in the TPCFs with
$D/R_{\rm vir}$ for blue galaxies ($0\sigma$), with {\vfifty} and
{\vninety} ($\sim 0.6$ and $\sim 1.65$, respectively) echoing this
result. Finally, in panel (d), we find no significant difference in
the TPCFs for red galaxies with $D/R_{\rm vir}$ ($2.1\sigma$);
however, we find that, while {\vfifty} is consistent within
uncertainties for the two subsamples ({\vfifty}$\sim 0.3$), the
{\vninety} is larger for the low $D/R_{\rm vir}$ subsample (0.95) than
the high $D/R_{\rm vir}$ subsample (0.75).

\begin{figure}[ht]
\includegraphics[width=\linewidth]{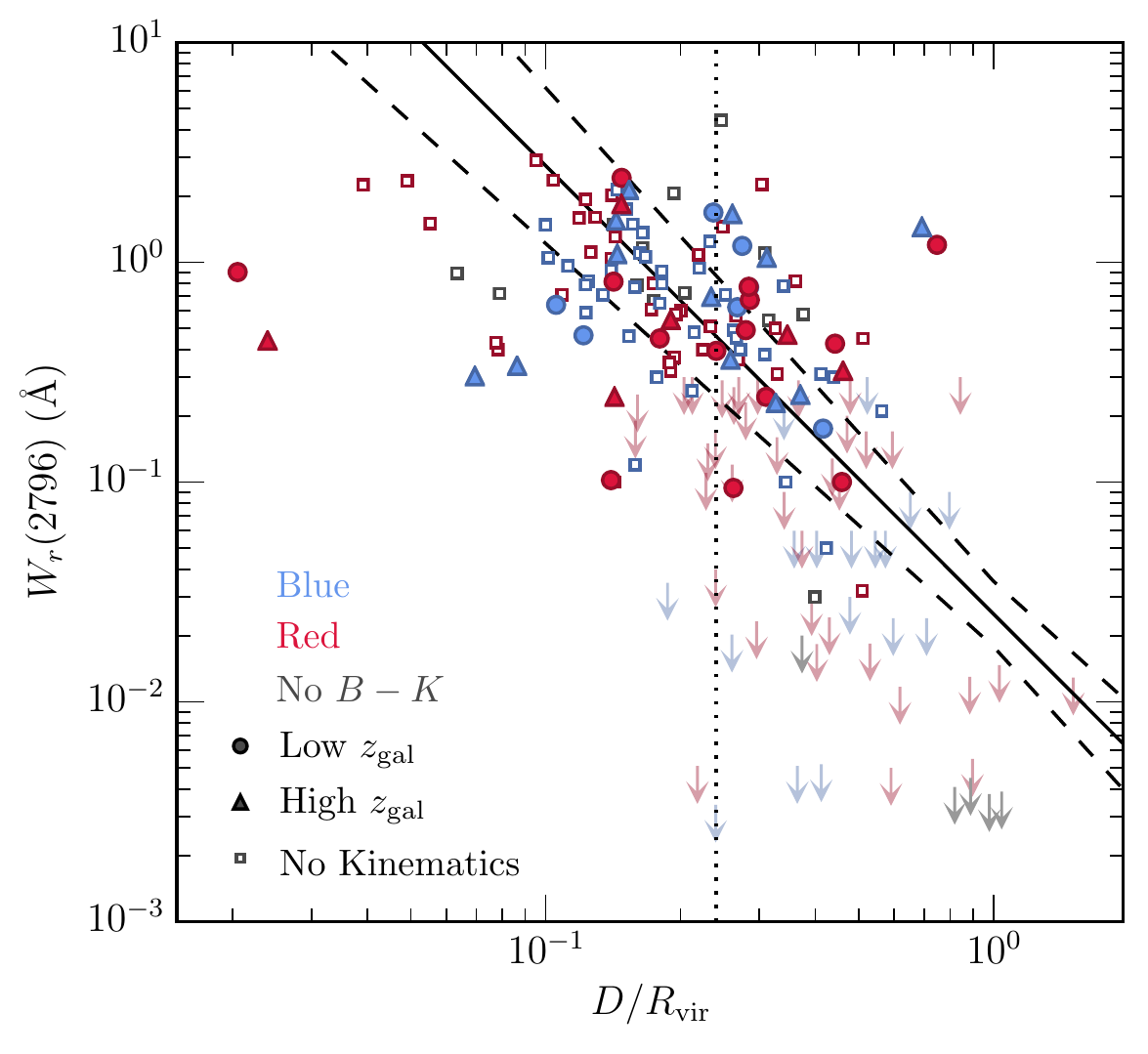}
\caption[]{{\MgII} equivalent width, $W_r(2796)$, versus $D/R_{\rm
    vir}$. Point colors indicate galaxy color sliced by $\langle B-K
  \rangle =1.4$. Solid filled points are those galaxies included in
  the sample presented here. Open points are the rest of the
  {\magiicat} \citep{magiicat1} absorber sample that do not have high
  resolution quasar spectra. Point types for the kinematics sample
  represent $z_{\rm gal}$, with circles for low $z_{\rm gal}$ and
  triangles for high $z_{\rm gal}$. Downward arrows are the
  nonabsorbing galaxies in the {\magiicat} sample. Solid and dashed
  lines are the fit to the data obtained by \citet{cwc-masses}. The
  vertical dotted line is the median $D/R_{\rm vir}$ of the kinematics
  sample, with $\langle D/R_{\rm vir} \rangle =0.24$. While the full
  {\magiicat} sample is anti-correlated to the $8.9\sigma$ level
  \citep{cwc-masses, magiicat3}, the kinematics sample (solid points)
  is not, with a significance of $0.8\sigma$.}
\label{fig:EWvsDR}
\end{figure}

{\it In summary}, the absorber velocity dispersion depends on
$D/R_{\rm vir}$ for red galaxies only, where the dispersions are
smaller at larger $D/R_{\rm vir}$, and this difference is present only
in the tails for the normalized TPCFs. At low $D/R_{\rm vir}$, the
velocity dispersions for absorption associated with blue and red
galaxies are comparable in the unnormalized TPCFs. In contrast, the
cloud column densities do not depend on whether the absorption is located
around blue or red galaxies, nor do they depend on $D/R_{\rm vir}$.

\subsection{Anti-correlation of $W_r(2796)$ and $D/R_{\rm vir}$}
\label{sec:WDR}

The results presented in the previous section (Section~\ref{sec:DR})
are puzzling given the anti-correlation between $W_r(2796)$ and
$D/R_{\rm vir}$ \citep{cwc-masses, magiicat3}. Since equivalent width
correlates with the number of clouds \citep{pb90, cvc03,
  evans-thesis}, the column densities, the velocity spreads, or both
should diminish with increasing $D/R_{\rm vir}$. Therefore, we
expected that the TPCFs and/or the cloud column densities would show a
dependence on $D/R_{\rm vir}$ regardless of color where the cloud
column densities and/or TPCF velocity dispersions would decrease with
increasing $D/R_{\rm vir}$. However, this is not the case. We found no
dependence of the TPCF velocity dispersions on $D/R_{\rm vir}$ for
blue galaxies, but the velocity dispersion for red galaxies is lower
at high $D/R_{\rm vir}$ (as might be expected) in the unnormalized
TPCFs. However, the red galaxy TPCF $D/R_{\rm vir}$ dependence
vanished when we normalized the pixel velocities by $V_{\rm
  circ}$. Additionally, the cloud column density distributions do not
differ with $D/R_{\rm vir}$ for both blue and red galaxies.

To better understand the sample examined here in the context of the
$W_r(2796)$--$D/R_{\rm vir}$ anti-correlation, we present
Figure~\ref{fig:EWvsDR} in which we plot the present sample of
galaxies as solid points, the rest of the {\magiicat} sample
\citep{magiicat1} absorbing galaxies as open points, and {\magiicat}
nonabsorbing galaxies as downward arrows as their absorption is only
known to a $3\sigma$ upper limit. This plot is similar to Figure~1(c)
in \citet{cwc-masses}, though point colors here represent galaxy
colors sliced by the median color, $\langle B-K \rangle =1.4$. The
solid and dashed lines are the fit to the data in
\citet{cwc-masses}. For reference we plot the median $D/R_{\rm vir}$
as a vertical dotted line.

As reported by \citet{cwc-masses}, the anti-correlation is significant
to the $8.9\sigma$ level for the full {\magiicat} sample using a
BHK-$\tau$ non-parametric rank correlation test to account for the
upper limits on absorption. Since we study only those systems with
detected absorption in high resolution quasar spectra, we ran a
Kendall-$\tau$ rank-correlation test between $W_r(2796)$ and $D/R_{\rm
  vir}$ for the present sample (filled points) and found an
anti-correlation that is significant to the $0.8\sigma$ level. Thus
the data presented here do not exhibit an anti-correlation between
$W_r(2796)$ and $D/R_{\rm vir}$ (however, if we examine {\it all}
absorbers in the {\magiicat} sample, we do find an anti-correlation
with $4.6\sigma$ significance). If we examine only blue galaxies, the
significance drops further to $0.2\sigma$ (for red galaxies the
significance remains at $0.8\sigma$). This result is then consistent
with no $D/R_{\rm vir}$ dependence for the cloud column densities
regardless of galaxy color and for the TPCFs of blue
galaxies. However, it does not explain the differences in the TPCFs of
absorbers associated with red galaxies in the unnormalized TPCFs.

We also examined the statistics on this anti-correlation for low and
high $z_{\rm gal}$ subsamples to determine if the anti-correlation was
affecting the redshift evolution results in the TPCFs. Point types in
Figure~\ref{fig:EWvsDR} represent $z_{\rm gal}$ subsamples, with
circles for low $z_{\rm gal}$ and triangles for high $z_{\rm
  gal}$. The Kendall-$\tau$ rank-correlation test resulted in an
insignificant anti-correlation between $W_r(2796)$ and $D/R_{\rm vir}$
for both the low $z_{\rm gal}$ ($1.1\sigma$) and the high $z_{\rm
  gal}$ ($0.1\sigma$) subsamples. 

To ensure that our kinematics sample is not unusual, we randomly drew
39 absorbers from the full {\magiicat} absorber sample for one million
realizations and ran the rank-correlation test each time. The fraction
of realizations in which the anti-correlation between $W_r(2796)$ and
$D/R_{\rm vir}$ is significant (i.e., the significance is greater than
$3\sigma$) is 25\%. For blue galaxies (19 absorbers) and similarly
with red galaxies (20 absorbers), this fraction drops to 4\%. Finding
no significant anti-correlation is not unusual. Given the history of
absorber-galaxy studies \citep[see e.g.,][for a list of
  references]{cwc-china, magiicat1}, this is not unexpected. With
larger numbers of absorber--galaxy pairs, the statistics on the
$W_r(2796)$--$D$ anti-correlation has steadily become more
significant. The main reason for the lack of an anti-correlation for a
given smaller sample is the large scatter in the relation.

\section{Discussion}
\label{sec:discussion}

By examining the kinematics and cloud column densities of {\MgII}
absorbers, we have observed redshift evolution in the CGM of red
galaxies where the velocity dispersions of absorbers decrease and the
cloud column densities increase with decreasing redshift. When
examining the kinematics as a function of $D/R_{\rm vir}$, we also
find a difference for red galaxies where the velocity dispersions
decrease with increasing $D/R_{\rm vir}$, though the cloud column
densities do not differ at low and high $D/R_{\rm vir}$. The radial
dependence in the velocity dispersions for red galaxies is removed
when we normalize the pixel velocities by $V_{\rm circ}$. Conversely,
we find no redshift or radial dependence of the velocity dispersions
and cloud column densities for blue galaxies. Compared to the red
galaxies, the blue galaxy velocity dispersions and cloud column
densities are larger than for red galaxies.

We found that red galaxies have smaller velocity dispersions than blue
galaxies overall; this is most obvious in
Figure~\ref{fig:bivariate}. Since blue (less massive) galaxies tend to
have a larger star formation rate than red (more massive) galaxies,
blue galaxies are more likely to experience outflows than red
galaxies. Thus, the large velocity dispersions in blue galaxies may
well be due to outflows induced by star formation which act to ``stir
up'' the {\MgII} absorbers, whereas a lack of outflows in red galaxies
likely causes the smaller velocity dispersions in red galaxy
TPCFs. This is consistent with previous works in which outflows were
invoked to explain the presence and properties of {\MgII} absorption
\citep[e.g.,][]{rubin-winds, rubin-winds14, bouche12, martin12,
  bordoloi14-model, bordoloi14, kacprzak14}.

A possible alternative explanation for the large TPCF velocity
dispersions for the blue galaxies is the presence of merging satellite
galaxies. Regardless of the host galaxy type, satellite galaxies
present such a small cross-section that they are unlikely to be a
significant source of {\MgII} absorption around host galaxies. Several
works have investigated this by comparing the estimated satellite
cross-sections from simulations to the observed incidence of
absorption and found that the satellite cross-sections are much lower
than the absorption incidence rate \citep[e.g.,][]{tumlinson13,
  gauthier10}. Thus, satellites are unlikely to explain the properties
of the {\MgII} absorbers we present here. For more discussion of
possible effects of satellite contributions to {\MgII} kinematics, see
Paper V of the {\magiicat} series \citep{magiicat5}.

That we find differences in the velocity dispersions and cloud column
densities for absorbers around red galaxies with redshift, but no such
evolution in blue galaxies, may suggest we are observing the
consequences of quenched star formation in red galaxies but ongoing
star formation in blue galaxies. Due to the ongoing star formation in
blue galaxies, the absorbers are likely to be involved in outflows,
accretion, and/or recycling at all redshifts, thus their velocity
dispersions remain large and their cloud column densities remain
unchanged. Outflows may continually replenish the CGM of disturbed,
large column density gas.

At high redshift in Figure~\ref{fig:BKzabs}(b), the absorber velocity
dispersions are similar regardless of galaxy color, indicating that
the red galaxies we observe in the high redshift subsample may have
undergone star formation driven outflows recently. At low redshift in
Figure~\ref{fig:BKzabs}(a), the TPCFs are narrower for red galaxies
than blue, possibly indicating that the outflows at higher redshift
have since shut off. This is also shown in the normalized TPCFs in
Figure~\ref{fig:BKzabsnorm}, where the high $z_{\rm gal}$, red galaxy
TPCF is more narrow than the blue galaxy TPCF, but this difference
increases for low $z_{\rm gal}$ subsamples. Since the cloud column
densities of red galaxies at high $z_{\rm gal}$ are smaller than those
for blue galaxies, some mechanism present only in red galaxies may
break the clouds into smaller column density clouds. At lower
redshift, the cloud column density distributions for blue and red
galaxies are comparable, but similar to the distribution for blue
galaxies at higher redshift.

The quenching of star formation may act to slowly reduce the velocity
dispersion of {\MgII} absorbers, but the quenching event initially
breaks the clouds into smaller column density clouds, which then
increase over time. This may be explained by a scenario in which star
forming galaxies have outflows driven by active star formation which
agitate and disperse the gas in the CGM to larger velocity
dispersions. Star formation is then shut off via an unknown quenching
mechanism (either AGN activity, intense star formation, a
galaxy--galaxy merger, etc.) which breaks the clouds into smaller
column densities. Over time, the absorbers are allowed to settle to
lower velocity dispersions, which then allows for the individual
clouds to ``re-condense'' to form larger column density
clouds. Alternatively, the cloud column densities may appear to
increase because as the velocity dispersion decreases, the gas builds
over a narrower velocity range, resulting in larger measured column
densities even if the individual clouds are physically separate with
unchanging column densities. Thus the CGM becomes quiescent.

Initially we found a dependence of the red galaxy TPCFs on $D/R_{\rm
  vir}$ where the velocity dispersion decreases with increasing
$D/R_{\rm vir}$, which would follow from the $W_r(2796)$--$D/R_{\rm
  vir}$ anti-correlation discussed in \citet{cwc-masses}. This result
and the lack of a dependence of the TPCFs for blue galaxies on
$D/R_{\rm vir}$ would suggest that outflows push disturbed material
out to large distances. Then when star formation is quenched, the
velocity dispersions of gas in the inner CGM remain large, but the
outer region of the CGM is the first to show signs of quenched star
formation in the form of decreasing velocity dispersions. However, by
investigating the $W_r(2796)$--$D/R_{\rm vir}$ relation in
Section~\ref{sec:WDR} we found that there is no anti-correlation for
the sample presented here with a $0.8\sigma$ significance (same for
red galaxies) nor blue galaxies whose significance drops to
$0.2\sigma$. This is likely due to the small sample size as well as
the fact that we focus only on absorbers, which have only a
$4.6\sigma$ anti-correlation, compared to the full sample that includes
nonabsorbers with $8.9\sigma$. Since the $W_r(2796)$--$D/R_{\rm vir}$
anti-correlation is not interfering with, nor contaminating the sample
presented here, the redshift evolution of the TPCFs and cloud column
densities is more strongly explained as being due to the quenching of
star formation rather than an underlying sample bias.

Since the most significant differences in the TPCFs with $D/R_{\rm
  vir}$ were for red galaxies and we have no $W_r(2796)$--$D/R_{\rm
  vir}$ anti-correlation for this sample, we examined the mass
distributions for each of the $B-K$ and $D/R_{\rm vir}$ subsample
combinations. We found that galaxies in the low $D/R_{\rm vir}$
subsamples tend to be slightly more massive than those in the high
$D/R_{\rm vir}$ subsamples, for both blue and red galaxies. Thus,
normalizing the pixel velocities in the TPCFs by $V_{\rm circ}$
removes the mass bias with velocity, where more massive galaxies can
have higher velocity gas. Doing this resulted in no differences in the
mass-normalized TPCFs with $D/R_{\rm vir}$ for both blue and red
galaxies. However, the TPCFs for red galaxies are still narrower than
for blue galaxies at all $D/R_{\rm vir}$. This may indicate that,
after accounting for the mass of the galaxy in both the size of the
CGM and velocities, the quenching of star formation in red galaxies
affects gas in the CGM at all $D/R_{\rm vir}$, at least out to
$D/R_{\rm vir}=0.75$.

While our findings here provide interesting details into the nature of
the CGM, a large body of work has shown that the characteristics of
the absorbing gas in the CGM depend on both the galaxy inclination and
whether the absorption is located along the galaxy projected major or
minor axes \citep[e.g.,][]{bordoloi11, bordoloi14, bouche12, ggk-sims,
  kcn12, lan14, rubin-winds14}. This is especially true for star
forming galaxies as outflowing gas tends to be found along the minor
axes \citep[e.g.,][]{bouche12, kacprzak14} while accretion tends to be
detected along the major axes \citep[e.g.,][]{ggk-sims,
  bouche13}. These trends have also been observed in simulations
\citep[e.g.,][]{stewart11, danovich12, danovich14}. In the data
presented here, there may be hints of a bimodality for the blue galaxy
subsample TPCFs and cloud column density distributions, which may be
due to these orientation effects. In fact, we examined these
orientation effects in a companion paper \citep[{\magiicat}
  V;][]{magiicat5} and found that the largest velocity dispersions
were associated with blue, face-on galaxies ($i<57^{\circ}$), and are
likely due to outflowing gas pointed towards the observer. The
smallest velocity dispersions were associated with red, face-on
galaxies, which may be due to a lack of outflows as the star formation
in red galaxies has been quenched. Similar velocity dispersions for
blue and red galaxies that are in edge-on inclinations
($i\geq57^{\circ}$), probed along the galaxy projected major axis,
indicated that the gas we observed was accreting/rotating around the
galaxies. These orientation results have larger significances in the
chi-squared results than the results we present here. Thus, the
orientation of the host galaxy may be more important than $z_{\rm
  gal}$ and $D/R_{\rm vir}$ in understanding the processes giving rise
to absorption in the CGM.

\section{Summary and Conclusions}
\label{sec:conclusions}

Using a subset of {\magiicat} galaxies \citep{magiicat1}, we examined
the kinematics of gas in the CGM as a function of galaxy color,
redshift, and virial radius-normalized impact parameter. Each galaxy
was spectroscopically identified to be located at the redshift of an
associated {\MgII} absorber in a high-resolution quasar spectrum
within a projected distance of $D=200$~kpc. Thus the galaxy sample is
an absorption-selected sample, and only those absorption regions with
$W_r(2796)\geq 0.04$~{\AA} were included in our analysis. Galaxy
virial radii and circular velocities were obtained using halo
abundance matching \citep{magiicat3} and were used to normalize out
any mass dependence with impact parameter and velocity.

Our main conclusions are as follows:

\begin{enumerate}[nolistsep]

\item We find no redshift evolution in the kinematics nor cloud column
  densities for absorbers hosted by blue galaxies. This is possibly
  due to ongoing star formation, which causes outflows that continue
  to agitate and disperse the absorbers to form large velocity
  dispersions. Outflows thus continually replenish the CGM with large
  column density, high velocity dispersion gas. This result is still
  true when we normalize the pixel velocities by $V_{\rm circ}$ to
  remove any mass dependence.

\item Conversely, we find redshift evolution in the kinematics for
  absorbers hosted by red galaxies. The quenching of star formation in
  red galaxies may shut off outflows, which then may prevent the CGM
  from being replenished with the large column density, high velocity
  dispersion gas seen in blue galaxies. Because of this, once star
  formation has been quenched, absorbers appear to relax into lower
  velocity dispersions. The quenching mechanism may act to reduce the
  cloud column densities initially, but the column densities increase
  towards lower redshifts. Due to the lower velocity dispersions at
  lower redshifts, the clouds may be able to ``re-condense'' into
  larger column density clouds, or appear to increase in column
  density due to a narrower velocity range over which the clouds are
  spread, regardless of the physical distance between clouds. This
  result also stands when we normalize the pixel velocities by $V_{\rm
    circ}$.

\item Despite an overall anti-correlation between {\MgII} equivalent
  width and $D/R_{\rm vir}$ reported in \citet{magiicat3}, the sample
  we present here (those absorbers for which we have high-resolution
  quasar spectra) does not follow a $W_r(2796)$--$D/R_{\rm vir}$
  anti-correlation. This strengthens the result that the kinematics
  and cloud column densities of red galaxies undergo redshift
  evolution since an underlying anti-correlation with $D/R_{\rm vir}$
  is not interfering with the result.

\item Neither the TPCFs nor the cloud column density distributions
  depend on $D/R_{\rm vir}$ when the absorber pixel velocities are
  normalized by $V_{\rm circ}$ (to remove any possible mass bias in
  the data). This is consistent with the lack of an anti-correlation
  between $W_r(2796)$ and $D/R_{\rm vir}$. Since the TPCFs for red
  galaxies are more narrow than for blue galaxies at all $D/R_{\rm
    vir}$, this suggests that quenching affects the CGM out to at
  least $D/R_{\rm vir}=0.75$.

\end{enumerate}

This work constitutes our first examination of the kinematics of
{\MgII} absorbers as a function of galaxy properties. Previous works
had examined the kinematics of the absorbers in a variety of ways but
had not connected their results to the host galaxy properties, at
least in a statistical fashion as is possible with the pixel-velocity
TPCFs. In future work and to further understand the kinematics of gas
in the CGM as a function of galaxy properties, we will shift the pixel
velocities to the galaxy systemic velocity. We will also examine the
gas kinematics as a function of star formation rate (SFR), specific
SFR, SFR density, and galaxy metallicity.

\acknowledgments 

We thank C.~Steidel and J.-R.~Gauthier for providing reduced
HIRES/Keck quasar spectra. This material is based upon work supported
by the National Science Foundation under Grant No. 1210200 (NSF East
Asia and Pacific Summer Institutes). N.M.N.~was also partially
supported through a NMSGC Graduate Fellowship and a Graduate Research
Enhancement Grant (GREG) sponsored by the Office of the Vice President
for Research at New Mexico State University. G.G.K.~acknowledges the
support of the Australian Research Council through the award of a
Future Fellowship (FT140100933). M.T.M.~thanks the Australian Research
Council for Discovery Project grant DP130100568 which supported this
work. This work is based in part on observations collected at the
European Organisation for Astronomical Research in the Southern
Hemisphere under ESO programs 076.A-0860, 69.A-0371, 68.A-0170,
075.A-0841, 65.O-0158, 072.A-0346, 074.A-0201, 67.A-0146, 278.A-5048,
076.A-0463, 67.A-0022, and 67.C-0157. Some of the data presented
herein were obtained at the W.M. Keck Observatory, which is operated
as a scientific partnership among the California Institute of
Technology, the University of California and the National Aeronautics
and Space Administration. The Observatory was made possible by the
generous financial support of the W.M. Keck Foundation.

\bibliographystyle{apj}
\bibliography{refs}

\begin{thebibliography}{57}
\expandafter\ifx\csname natexlab\endcsname\relax\def\natexlab#1{#1}\fi

\bibitem[{{Barton} \& {Cooke}(2009)}]{bc09}
{Barton}, E.~J., \& {Cooke}, J. 2009, \aj, 138, 1817

\bibitem[{{Bergeron} \& {Boiss{\'e}}(1991)}]{bb91}
{Bergeron}, J., \& {Boiss{\'e}}, P. 1991, \aap, 243, 344

\bibitem[{{Bordoloi} {et~al.}(2014{\natexlab{a}}){Bordoloi}, {Lilly},
  {Kacprzak}, \& {Churchill}}]{bordoloi14-model}
{Bordoloi}, R., {Lilly}, S.~J., {Kacprzak}, G.~G., \& {Churchill}, C.~W.
  2014{\natexlab{a}}, \apj, 784, 108

\bibitem[{{Bordoloi} {et~al.}(2011){Bordoloi}, {Lilly}, {Knobel}, {Bolzonella},
  {Kampczyk}, {Carollo}, {Iovino}, {Zucca}, {Contini}, {Kneib}, {Le Fevre},
  {Mainieri}, {Renzini}, {Scodeggio}, {Zamorani}, {Balestra}, {Bardelli},
  {Bongiorno}, {Caputi}, {Cucciati}, {de la Torre}, {de Ravel}, {Garilli},
  {Kova{\v c}}, {Lamareille}, {Le Borgne}, {Le Brun}, {Maier}, {Mignoli},
  {Pello}, {Peng}, {Perez Montero}, {Presotto}, {Scarlata}, {Silverman},
  {Tanaka}, {Tasca}, {Tresse}, {Vergani}, {Barnes}, {Cappi}, {Cimatti},
  {Coppa}, {Diener}, {Franzetti}, {Koekemoer}, {L{\'o}pez-Sanjuan},
  {McCracken}, {Moresco}, {Nair}, {Oesch}, {Pozzetti}, \&
  {Welikala}}]{bordoloi11}
{Bordoloi}, R., {Lilly}, S.~J., {Knobel}, C., {et~al.} 2011, \apj, 743, 10

\bibitem[{{Bordoloi} {et~al.}(2014{\natexlab{b}}){Bordoloi}, {Lilly},
  {Hardmeier}, {Contini}, {Kneib}, {Le Fevre}, {Mainieri}, {Renzini},
  {Scodeggio}, {Zamorani}, {Bardelli}, {Bolzonella}, {Bongiorno}, {Caputi},
  {Carollo}, {Cucciati}, {de la Torre}, {de Ravel}, {Garilli}, {Iovino},
  {Kampczyk}, {Kova{\v c}}, {Knobel}, {Lamareille}, {Le Borgne}, {Le Brun},
  {Maier}, {Mignoli}, {Oesch}, {Pello}, {Peng}, {Perez Montero}, {Presotto},
  {Silverman}, {Tanaka}, {Tasca}, {Tresse}, {Vergani}, {Zucca}, {Cappi},
  {Cimatti}, {Coppa}, {Franzetti}, {Koekemoer}, {Moresco}, {Nair}, \&
  {Pozzetti}}]{bordoloi14}
{Bordoloi}, R., {Lilly}, S.~J., {Hardmeier}, E., {et~al.} 2014{\natexlab{b}},
  \apj, 794, 130

\bibitem[{{Bouch{\'e}} {et~al.}(2012){Bouch{\'e}}, {Hohensee}, {Vargas},
  {Kacprzak}, {Martin}, {Cooke}, \& {Churchill}}]{bouche12}
{Bouch{\'e}}, N., {Hohensee}, W., {Vargas}, R., {et~al.} 2012, \mnras, 426, 801

\bibitem[{{Bouch{\'e}} {et~al.}(2013){Bouch{\'e}}, {Murphy}, {Kacprzak},
  {P{\'e}roux}, {Contini}, {Martin}, \& {Dessauges-Zavadsky}}]{bouche13}
{Bouch{\'e}}, N., {Murphy}, M.~T., {Kacprzak}, G.~G., {et~al.} 2013, Science,
  341, 50

\bibitem[{{Chen} {et~al.}(2010){Chen}, {Helsby}, {Gauthier}, {Shectman},
  {Thompson}, \& {Tinker}}]{chen10a}
{Chen}, H.-W., {Helsby}, J.~E., {Gauthier}, J.-R., {et~al.} 2010, \apj, 714,
  1521

\bibitem[{{Churchill}(1997)}]{cwc-thesis}
{Churchill}, C.~W. 1997, PhD thesis, {University of California, Santa Cruz}

\bibitem[{{Churchill} {et~al.}(2005){Churchill}, {Kacprzak}, \&
  {Steidel}}]{cwc-china}
{Churchill}, C.~W., {Kacprzak}, G.~G., \& {Steidel}, C.~C. 2005, in IAU Colloq.
  199: Probing Galaxies through Quasar Absorption Lines, ed. P.~{Williams},
  C.-G. {Shu}, \& B.~{Menard}, 24--41

\bibitem[{{Churchill} {et~al.}(2013{\natexlab{a}}){Churchill}, {Nielsen},
  {Kacprzak}, \& {Trujillo-Gomez}}]{cwc-masses}
{Churchill}, C.~W., {Nielsen}, N.~M., {Kacprzak}, G.~G., \& {Trujillo-Gomez},
  S. 2013{\natexlab{a}}, \apjl, 763, L42

\bibitem[{{Churchill} {et~al.}(2013{\natexlab{b}}){Churchill},
  {Trujillo-Gomez}, {Nielsen}, \& {Kacprzak}}]{magiicat3}
{Churchill}, C.~W., {Trujillo-Gomez}, S., {Nielsen}, N.~M., \& {Kacprzak},
  G.~G. 2013{\natexlab{b}}, \apj, 779, 87

\bibitem[{{Churchill} \& {Vogt}(2001)}]{cv01}
{Churchill}, C.~W., \& {Vogt}, S.~S. 2001, \aj, 122, 679

\bibitem[{{Churchill} {et~al.}(2003){Churchill}, {Vogt}, \& {Charlton}}]{cvc03}
{Churchill}, C.~W., {Vogt}, S.~S., \& {Charlton}, J.~C. 2003, \aj, 125, 98

\bibitem[{{Crighton} {et~al.}(2013){Crighton}, {Hennawi}, \&
  {Prochaska}}]{crighton13}
{Crighton}, N.~H.~M., {Hennawi}, J.~F., \& {Prochaska}, J.~X. 2013, \apjl, 776,
  L18

\bibitem[{{Danovich} {et~al.}(2015){Danovich}, {Dekel}, {Hahn}, {Ceverino}, \&
  {Primack}}]{danovich14}
{Danovich}, M., {Dekel}, A., {Hahn}, O., {Ceverino}, D., \& {Primack}, J. 2015,
  \mnras, 449, 2087

\bibitem[{{Danovich} {et~al.}(2012){Danovich}, {Dekel}, {Hahn}, \&
  {Teyssier}}]{danovich12}
{Danovich}, M., {Dekel}, A., {Hahn}, O., \& {Teyssier}, R. 2012, \mnras, 422,
  1732

\bibitem[{{Dekker} {et~al.}(2000){Dekker}, {D'Odorico}, {Kaufer}, {Delabre}, \&
  {Kotzlowski}}]{dekker-uves}
{Dekker}, H., {D'Odorico}, S., {Kaufer}, A., {Delabre}, B., \& {Kotzlowski}, H.
  2000, in SPIE Conference Series, Vol. 4008, Optical and IR Telescope
  Instrumentation and Detectors, ed. M.~{Iye} \& A.~F. {Moorwood}, 534--545

\bibitem[{{Evans}(2011)}]{evans-thesis}
{Evans}, J.~L. 2011, PhD thesis, {New Mexico State University}

\bibitem[{{Fox} {et~al.}(2015){Fox}, {Bordoloi}, {Savage}, {Lockman},
  {Jenkins}, {Wakker}, {Bland-Hawthorn}, {Hernandez}, {Kim}, {Benjamin},
  {Bowen}, \& {Tumlinson}}]{fox15}
{Fox}, A.~J., {Bordoloi}, R., {Savage}, B.~D., {et~al.} 2015, \apjl, 799, L7

\bibitem[{{Gauthier} {et~al.}(2010){Gauthier}, {Chen}, \&
  {Tinker}}]{gauthier10}
{Gauthier}, J.-R., {Chen}, H.-W., \& {Tinker}, J.~L. 2010, \apj, 716, 1263

\bibitem[{{Guillemin} \& {Bergeron}(1997)}]{guillemin97}
{Guillemin}, P., \& {Bergeron}, J. 1997, \aap, 328, 499

\bibitem[{{Hopkins} \& {Beacom}(2006)}]{hopkins06}
{Hopkins}, A.~M., \& {Beacom}, J.~F. 2006, \apj, 651, 142

\bibitem[{{Kacprzak} {et~al.}(2010){Kacprzak}, {Churchill}, {Ceverino},
  {Steidel}, {Klypin}, \& {Murphy}}]{ggk-sims}
{Kacprzak}, G.~G., {Churchill}, C.~W., {Ceverino}, D., {et~al.} 2010, \apj,
  711, 533

\bibitem[{{Kacprzak} {et~al.}(2011){Kacprzak}, {Churchill}, {Evans}, {Murphy},
  \& {Steidel}}]{kcems11}
{Kacprzak}, G.~G., {Churchill}, C.~W., {Evans}, J.~L., {Murphy}, M.~T., \&
  {Steidel}, C.~C. 2011, \mnras, 416, 3118

\bibitem[{{Kacprzak} {et~al.}(2012{\natexlab{a}}){Kacprzak}, {Churchill}, \&
  {Nielsen}}]{kcn12}
{Kacprzak}, G.~G., {Churchill}, C.~W., \& {Nielsen}, N.~M. 2012{\natexlab{a}},
  \apjl, 760, L7

\bibitem[{{Kacprzak} {et~al.}(2008){Kacprzak}, {Churchill}, {Steidel}, \&
  {Murphy}}]{ggk08}
{Kacprzak}, G.~G., {Churchill}, C.~W., {Steidel}, C.~C., \& {Murphy}, M.~T.
  2008, \aj, 135, 922

\bibitem[{{Kacprzak} {et~al.}(2012{\natexlab{b}}){Kacprzak}, {Churchill},
  {Steidel}, {Spitler}, \& {Holtzman}}]{ggk1317}
{Kacprzak}, G.~G., {Churchill}, C.~W., {Steidel}, C.~C., {Spitler}, L.~R., \&
  {Holtzman}, J.~A. 2012{\natexlab{b}}, \mnras, 427, 3029

\bibitem[{{Kacprzak} {et~al.}(2014){Kacprzak}, {Martin}, {Bouch{\'e}},
  {Churchill}, {Cooke}, {LeReun}, {Schroetter}, {Ho}, \& {Klimek}}]{kacprzak14}
{Kacprzak}, G.~G., {Martin}, C.~L., {Bouch{\'e}}, N., {et~al.} 2014, \apjl,
  792, L12

\bibitem[{{Lan} {et~al.}(2014){Lan}, {M{\'e}nard}, \& {Zhu}}]{lan14}
{Lan}, T.-W., {M{\'e}nard}, B., \& {Zhu}, G. 2014, \apj, 795, 31

\bibitem[{{Lanzetta} \& {Bowen}(1990)}]{lanzetta90}
{Lanzetta}, K.~M., \& {Bowen}, D. 1990, \apj, 357, 321

\bibitem[{{Lilly} {et~al.}(2013){Lilly}, {Carollo}, {Pipino}, {Renzini}, \&
  {Peng}}]{lilly-bathtub}
{Lilly}, S.~J., {Carollo}, C.~M., {Pipino}, A., {Renzini}, A., \& {Peng}, Y.
  2013, \apj, 772, 119

\bibitem[{{Martin} {et~al.}(2012){Martin}, {Shapley}, {Coil}, {Kornei},
  {Bundy}, {Weiner}, {Noeske}, \& {Schiminovich}}]{martin12}
{Martin}, C.~L., {Shapley}, A.~E., {Coil}, A.~L., {et~al.} 2012, \apj, 760, 127

\bibitem[{{Matejek} \& {Simcoe}(2012)}]{matejek12}
{Matejek}, M.~S., \& {Simcoe}, R.~A. 2012, \apj, 761, 112

\bibitem[{{Mathes} {et~al.}(2014){Mathes}, {Churchill}, {Kacprzak}, {Nielsen},
  {Trujillo-Gomez}, {Charlton}, \& {Muzahid}}]{mathes14}
{Mathes}, N.~L., {Churchill}, C.~W., {Kacprzak}, G.~G., {et~al.} 2014, \apj,
  792, 128

\bibitem[{{M{\'e}nard} {et~al.}(2011){M{\'e}nard}, {Wild}, {Nestor}, {Quider},
  {Zibetti}, {Rao}, \& {Turnshek}}]{menard11}
{M{\'e}nard}, B., {Wild}, V., {Nestor}, D., {et~al.} 2011, \mnras, 417, 801

\bibitem[{{Nielsen} {et~al.}(2013{\natexlab{a}}){Nielsen}, {Churchill}, \&
  {Kacprzak}}]{magiicat2}
{Nielsen}, N.~M., {Churchill}, C.~W., \& {Kacprzak}, G.~G. 2013{\natexlab{a}},
  \apj, 776, 115

\bibitem[{{Nielsen} {et~al.}(2013{\natexlab{b}}){Nielsen}, {Churchill},
  {Kacprzak}, \& {Murphy}}]{magiicat1}
{Nielsen}, N.~M., {Churchill}, C.~W., {Kacprzak}, G.~G., \& {Murphy}, M.~T.
  2013{\natexlab{b}}, \apj, 776, 114

\bibitem[{{Nielsen} {et~al.}(2015){Nielsen}, {Churchill}, {Kacprzak}, {Murphy},
  \& {Evans}}]{magiicat5}
{Nielsen}, N.~M., {Churchill}, C.~W., {Kacprzak}, G.~G., {Murphy}, M.~T., \&
  {Evans}, J.~L. 2015, \apj, 812, 83

\bibitem[{{Oppenheimer} \& {Dav{\'e}}(2008)}]{oppenheimer08}
{Oppenheimer}, B.~D., \& {Dav{\'e}}, R. 2008, \mnras, 387, 577

\bibitem[{{Petitjean} \& {Bergeron}(1990)}]{pb90}
{Petitjean}, P., \& {Bergeron}, J. 1990, \aap, 231, 309

\bibitem[{{Ribaudo} {et~al.}(2011){Ribaudo}, {Lehner}, {Howk}, {Werk}, {Tripp},
  {Prochaska}, {Meiring}, \& {Tumlinson}}]{ribaudo11}
{Ribaudo}, J., {Lehner}, N., {Howk}, J.~C., {et~al.} 2011, \apj, 743, 207

\bibitem[{{Rubin} {et~al.}(2012){Rubin}, {Prochaska}, {Koo}, \&
  {Phillips}}]{rubin-accretion}
{Rubin}, K.~H.~R., {Prochaska}, J.~X., {Koo}, D.~C., \& {Phillips}, A.~C. 2012,
  \apjl, 747, L26

\bibitem[{{Rubin} {et~al.}(2014){Rubin}, {Prochaska}, {Koo}, {Phillips},
  {Martin}, \& {Winstrom}}]{rubin-winds14}
{Rubin}, K.~H.~R., {Prochaska}, J.~X., {Koo}, D.~C., {et~al.} 2014, \apj, 794,
  156

\bibitem[{{Rubin} {et~al.}(2010){Rubin}, {Weiner}, {Koo}, {Martin},
  {Prochaska}, {Coil}, \& {Newman}}]{rubin-winds}
{Rubin}, K.~H.~R., {Weiner}, B.~J., {Koo}, D.~C., {et~al.} 2010, \apj, 719,
  1503

\bibitem[{{Sargent} {et~al.}(1988){Sargent}, {Steidel}, \&
  {Boksenberg}}]{ssb88}
{Sargent}, W.~L.~W., {Steidel}, C.~C., \& {Boksenberg}, A. 1988, \apj, 334, 22

\bibitem[{{Schneider} {et~al.}(1993){Schneider}, {Hartig}, {Jannuzi},
  {Kirhakos}, {Saxe}, {Weymann}, {Bahcall}, {Bergeron}, {Boksenberg},
  {Sargent}, {Savage}, {Turnshek}, \& {Wolfe}}]{schneider93}
{Schneider}, D.~P., {Hartig}, G.~F., {Jannuzi}, B.~T., {et~al.} 1993, \apjs,
  87, 45

\bibitem[{{Steidel} {et~al.}(1997){Steidel}, {Dickinson}, {Meyer},
  {Adelberger}, \& {Sembach}}]{steidel97}
{Steidel}, C.~C., {Dickinson}, M., {Meyer}, D.~M., {Adelberger}, K.~L., \&
  {Sembach}, K.~R. 1997, \apj, 480, 568

\bibitem[{{Steidel} {et~al.}(1994){Steidel}, {Dickinson}, \& {Persson}}]{sdp94}
{Steidel}, C.~C., {Dickinson}, M., \& {Persson}, S.~E. 1994, \apjl, 437, L75

\bibitem[{{Steidel} {et~al.}(2002){Steidel}, {Kollmeier}, {Shapley},
  {Churchill}, {Dickinson}, \& {Pettini}}]{steidel02}
{Steidel}, C.~C., {Kollmeier}, J.~A., {Shapley}, A.~E., {et~al.} 2002, \apj,
  570, 526

\bibitem[{{Stewart} {et~al.}(2011){Stewart}, {Kaufmann}, {Bullock}, {Barton},
  {Maller}, {Diemand}, \& {Wadsley}}]{stewart11}
{Stewart}, K.~R., {Kaufmann}, T., {Bullock}, J.~S., {et~al.} 2011, \apj, 738,
  39

\bibitem[{{Thom} {et~al.}(2011){Thom}, {Werk}, {Tumlinson}, {Prochaska},
  {Meiring}, {Tripp}, \& {Sembach}}]{thom11}
{Thom}, C., {Werk}, J.~K., {Tumlinson}, J., {et~al.} 2011, \apj, 736, 1

\bibitem[{{Tumlinson} {et~al.}(2011){Tumlinson}, {Thom}, {Werk}, {Prochaska},
  {Tripp}, {Weinberg}, {Peeples}, {O'Meara}, {Oppenheimer}, {Meiring}, {Katz},
  {Dav{\'e}}, {Ford}, \& {Sembach}}]{tumlinson11}
{Tumlinson}, J., {Thom}, C., {Werk}, J.~K., {et~al.} 2011, Science, 334, 948

\bibitem[{{Tumlinson} {et~al.}(2013){Tumlinson}, {Thom}, {Werk}, {Prochaska},
  {Tripp}, {Katz}, {Dav{\'e}}, {Oppenheimer}, {Meiring}, {Ford}, {O'Meara},
  {Peeples}, {Sembach}, \& {Weinberg}}]{tumlinson13}
---. 2013, \apj, 777, 59

\bibitem[{{Vogt} {et~al.}(1994){Vogt}, {Allen}, {Bigelow}, {Bresee}, {Brown},
  {Cantrall}, {Conrad}, {Couture}, {Delaney}, {Epps}, {Hilyard}, {Hilyard},
  {Horn}, {Jern}, {Kanto}, {Keane}, {Kibrick}, {Lewis}, {Osborne},
  {Pardeilhan}, {Pfister}, {Ricketts}, {Robinson}, {Stover}, {Tucker}, {Ward},
  \& {Wei}}]{vogt-hires}
{Vogt}, S.~S., {Allen}, S.~L., {Bigelow}, B.~C., {et~al.} 1994, in SPIE
  Conference Series, Vol. 2198, Instrumentation in Astronomy VIII, ed. D.~L.
  {Crawford} \& E.~R. {Craine}, 362

\bibitem[{{Weiner} {et~al.}(2009){Weiner}, {Coil}, {Prochaska}, {Newman},
  {Cooper}, {Bundy}, {Conselice}, {Dutton}, {Faber}, {Koo}, {Lotz}, {Rieke}, \&
  {Rubin}}]{weiner09}
{Weiner}, B.~J., {Coil}, A.~L., {Prochaska}, J.~X., {et~al.} 2009, \apj, 692,
  187

\bibitem[{{Werk} {et~al.}(2013){Werk}, {Prochaska}, {Thom}, {Tumlinson},
  {Tripp}, {O'Meara}, \& {Peeples}}]{werk13}
{Werk}, J.~K., {Prochaska}, J.~X., {Thom}, C., {et~al.} 2013, \apjs, 204, 17

\end{thebibliography}

\end{document}